
\hoffset = .75   in
\voffset = .75   in
\centerline{\bf Van Hove Excitons and High-T$_c$ Superconductivity: VIIIC}
\centerline{\bf Dynamic Jahn-Teller Effects vs Spin-Orbit Coupling in the LTO
Phase of La$_{2-x}$Sr$_x$CuO$_4$}
\smallskip
\centerline{R.S. Markiewicz}
\centerline{Physics Department and Barnett Institute, Northeastern University,
Boston, MA 02115}
\bigskip
\par\noindent
Running Title: {\bf Dynamic JT in LSCO}
\bigskip
\par
The possible role of the van Hove singularity (vHs) in stabilizing the
low-temperature orthorhombic (LTO) phase transition in
La$_{2-x}$\-Sr$_x$\-CuO$_
4$ (LSCO) is discussed.  It is found that the vHs can drive a structural
distortion in two different ways, either due to spin-orbit coupling or to
dynamic Jahn-Teller (JT) effects.  This paper discusses the latter effect in
some detail.
It is shown that a model Hamiltonian introduced earlier to describe the
coupled electron -- octahedral tilt motions (`cageons') has a series of
phase transitions, from a high-temperature disordered JT phase (similar to
the high-temperature tetragonal phase of LSCO) to an intermediate temperature
dynamic JT phase, of average orthorhombic symmetry (the LTO phase) to a low
temperature static JT phase (the low temperature tetragonal phase).  For some
parameter values, the static JT phase is absent.
\bigskip
\noindent{\bf 1. Introduction}
\bigskip\par
The high-T$_c$ superconductors La$_{2-x}$Ba$_x$CuO$_4$ (LBCO) and
La$_{2-x}$Sr$_
x$CuO$_4$ (LSCO) have structural phase transitions from a high-temperature
tetragonal (HTT) to a low-temperature orthorhombic (LTO) to (in doped LBCO)
a low-temperature tetragonal (LTT) phase[1].  There is a clear association of
the LTT phase with electronic properties -- the transition causes a decrease in
the Hall density and interferes destructively with superconductivity[2].
Moreover, the transition is complete only at a fixed hole density, $x\sim
0.125$[3].  These features can be understood[4] in terms of the LTT phase
splitting the degeneracy of the two van Hove singularities (vHs's)[5-7], at the
$X$- and $Y$-points of the (HTT) Brillouin zone.
\par
The role of electron-phonon coupling in the LTO phase is less clear, since
transport properties are only weakly affected by the HTT-LTO transition.
While a uniform LTO phase could be stabilized by
purely non-electronic mechanisms, this does not explain why high-T$_c$
superconductivity appears to occur only in the orthorhombic phase[8].
It had early been suggested that the LTO phase was a charge-density wave (CDW)
phase, associated with the vHs, but Pouget, et al.[9] suggested that this
interpretation was untenable, since the LTO transition does not split the
degeneracy of the two vHs's.  In the present paper, I show that the analysis
of Pouget, et al., is incomplete: the vHs can actually drive {\it two}
transitions of {\it macroscopic} LTO symmetry, to either a CDW-like phase or
to a spin-density wave (SDW)-like phase.
\par
I have recently suggested[10,11] that {\it both} LTT and LTO phases are
manifestations of a novel form of band Jahn-Teller (JT) effect[12-14], in which
the degenerate electronic states are associated with the two vHs's.
Thus, the LTT phase involves an essentially static band JT distortion,
splitting
the degeneracy of the two vHs's.  The HTT and LTO phases can then both be
interpreted as dynamic Jahn-Teller phases, involving
tunneling between the X- and Y- point JT distortions of the LTT phase.  In Ref.
[10], a mean field calculation was made, approximating the dynamic JT effect by
`valence bond density waves', a coherent superposition of two charge density
waves (CDW's).  The resulting phase diagram reproduced the HTT $\rightarrow$
LTO $\rightarrow$ superconducting phases as a function of hole doping in LSCO.
Ref. [11] introduced a model Hamiltonian to describe the (nonlinear)
electron-phonon interaction, based on similar calculations for the A15
compounds[15,16], and applied it to an analysis of the static JT effect.  The
present paper extends these calculations to include the dynamic JT effect.  For
a single cell, the combined electron-tilt phonon (`cageon') problem can be
reduced to the problem of a particle moving among four potential wells, and a
solution can be found in terms of Mathieu's functions.  The ground states are
linear combinations of tilted octahedra.  Adding intercell coupling leads to
phase transitions with a net macroscopic average tilt, and the accompanying
static strains.  A dynamic LTO phase can be stabilized over a considerable
temperature range by entropic effects, and the sequence HTT$\rightarrow$LTO$
\rightarrow$LTT is naturally reproduced.  A sequence of transitions is found,
from a high temperature disordered JT phase (equivalent to the HTT phase of
LSCO) to an intermediate dynamic JT phase, with equal tilts along the
orthogonal
x and y axes (an LTO phase) to a low temperature static JT phase (LTT phase).
The transition to the LTO phase is second order, while that between LTO and LTT
is first order.  Within a certain parameter range, the LTT phase does not
occur,
leaving the dynamic JT phase as the stable low-T phase.
\par
The dynamic JT phase offers an interpretation of the LTO phase as having only
{\it macroscopic average} orthorhombic symmetry, with local dynamic disorder.
The question of whether the vHs can induce a transition into a {\it uniform}
LTO phase is also reanalyzed, on the basis of group theory.  It is found that
umklapp scattering {\it can} split the vHs degeneracy, but only in the presence
of spin-orbit coupling.  Thus there are two vHs-based mechanisms of driving the
LTO transition, in competition with each other.  While considerably more
complicated, these transitions are reminiscent of the competition between
CDW and SDW transitions in the theory of nesting instabilities in
lower-dimensional metals.
\par
The paper is organized as follows.  Section 2 reviews the proposed role of band
JT effects in the cuprates, and their relation to ferroelectric perovskites.
Section 3 introduces a modified form of the electron-tilt-strain coupled
Hamiltonian of Ref. [11], including umklapp scattering terms.  Section 4 and
Appendix II show
how the acoustic strains can be formally eliminated, leading to an effective
electron-electron interaction.  Section 5 and Appendix III carefully analyze
the problem of structural phase transitions in the {\it uniform} LTO phase.
It is shown that, whereas umklapp scattering can in principle lead to a density
wave instability, the transition is symmetry forbidden for the LTO space group,
unless spin-orbit interaction is included.  Section 6 analyzes the dynamic JT
state, reducing the Hamiltonian to an intracell plus near neighbor coupling
form, and shows the relation between the present Hamiltonian and previous
JT calculations.  Section 7 presents the calculation of the dynamic JT effects
at a mean-field level, including phase diagrams of the transitions from
HTT$\rightarrow$LTO$\rightarrow$LTT.  A discussion is given in Section 8, while
the interpretation of the `cageon' in terms of polarons or solitons is briefly
discussed in Section 9.  A number of Appendices provide details of
calculations:  a renormalization of the band structure to reveal separate
subbands associated with each vHs (Appendix I); a group theoretical analysis of
the uniform LTO phase (Appendix III); a pseudospin
approximation of the electronic operators (Appendix IV); and a discussion of
the
solutions of the four-well Mathieu equation (Appendix V).
\bigskip
\noindent{\bf 2. First-Order Band JT Effect}
\smallskip\par
The JT theorem states that any orbital electronic degeneracy in a molecule is
unstable: there always exists some structural distortion which lowers the
energy of the molecule by lifting the degeneracy.  A similar effect
can arise in a solid, either for a local impurity or for a collective JT effect
in the crystal as a whole.  Many perovskites and related structures have
a phase transition to either a ferroelectric or antiferrodistortive phase,
which
can be interpreted as a collective JT effect[13].
\par
In the cuprates, the structural transitions are complicated, involving tilts
of the oxygen octahedra (optical phonons), static strains and acoustic phonons,
as well as electron-phonon coupling.  Moreover, the coupling of the optical
phonons to both the strains and the electrons is nonlinear.  A model for
these interactions was introduced in Ref. [11].  There, it was suggested that
both the LTO and LTT phase transitions were driven by the diverging electronic
susceptibility associated with the vHs.  Indeed, these transitions can be
considered as a generalization to two dimensitions of the Peierls instability.
\par
The LTT phase transition can be thought of as a vHs-JT transition.  The
presence
of two vHs's, at the X and Y points of the Brillouin zone, provides the
electronic degeneracy, while the tilting of the octahedra split this
degeneracy,
as in a traditional (band) JT effect.  Interpretation of the LTO phase is
more complicated, since the associated octahedral tilting leaves the vHs's
degenerate, essentially because the two in-plane Cu-O distances are equal in
the
LTO phase[9].  However, this rules out only the simplest model of the CDW/SDW
phase.  In the present paper, I show that there remain a large number of
potential ground states, stabilized by (partial) splitting of the vHs
degeneracy.  I suggest that two of these states may be actually realized in
LSCO: a static, SDW-like phase near half filling (Section 5), and a dynamic,
CDW-like phase in the doped material (Section 7).
\par
The difference between the two transitions can be understood from Fig. 1, which
illustrates two different means of splitting the vHs peak in the dos.
This Figure illustrates the Fermi surfaces corresponding to the LTT phase (Fig.
1a) and to a possible LTO phase (Fig. 1b).  The Fermi surfaces
at the vHs have been distorted from square by including a finite oxygen-oxygen
hopping energy, $t_{OO}$ in the dispersion relations.
\par
Figure 1a shows Fermi surfaces of the LTT phase for two different dopings
-- the Fermi level coincides with the $X$-point vHs at one doping, and with the
$Y$-point vHs at the other.  In the LTT phase, the two Cu-O distances are no
longer degenerate, so the Cu-O hopping parameter becomes anisotropic, splitting
the vHs degeneracy[4].  (Due to the unusual layering, this compound has a
complicated c-axis dispersion: the roles of $X$-point and $Y$-point vHs's are
interchanged on every other layer.  It might be thought that such a large
c-axis dispersion would wash out the effects of the vHs, but in fact, it is a
consequence of the vHs splitting in the LTT phase.)  In the dynamic model of
the LTO phase, the symmetry is microscopically LTT, thereby locally splitting
the vHs degeneracy, but with a macroscopic orthorhombic strain.
\par
This is to be contrasted with the situation in the uniform LTO phase, Fig. 1b.
The LTO transition doubles the unit cell volume, thereby reducing the
Brillouin zone volume by the same factor, and introducing a two-fold
degeneracy in all bands.  Unklapp scattering can in principle lift the
degeneracy of the two bands, so that their respective vHs's coincide with the
Fermi level at two different dopings, as illustrated in Fig. 1b.  Note that in
Fig. 1a, the Fermi surface of a single band is illustrated at two different
dopings, whereas Fig. 1b shows the Fermi surfaces of two different bands at a
single doping level.  Note further that, within either band in Fig. 1b, the
two vHs's remain degenerate[9].  The splitting of Fig. 1b can be used, for
instance, to describe interlayer coupling in the cuprates.  In this case, the
two `bands' would correspond to the Fermi surfaces at $k_z=0,\pi /c$, which are
degenerate when interlayer hopping is neglected.  However, interlayer coupling
is weak in these materials, and for simplicity it will be neglected.
\par
Of more interest is the possibility of introducing a gap by coherently coupling
excitations between the two vHs's.  Such in-plane coupling was introduced by
Schulz[17], and is very similar to that introduced in the flux phase[18]
of a doped antiferromagnet.  However, a group theoretical analysis rules out
such a possibility for a CDW-like excitation: there is a two-fold degeneracy
of electronic levels on the Brillouin zone face, which can only be split by
spin-orbit coupling.  This suggests that vHs splitting can lead to a uniform
LTO phase only in the presence of antiferromagnetic order -- i.e., close to
half
filling.  Away from half filling, the magnetic state becomes inhomogeneous[19].
Hence, in the doping range associated with superconductivity, there is
competition between a disordered magnetic state and a dynamic JT state.  Both
phases will be described in greater detail below.
\par
It seems likely that dynamic effects are important in the higher-T$_c$
phases, and that the sequence HTT$\rightarrow$LTO$\rightarrow$LTT may parallel
the sequence of structural transitions in the perovskite, BaTiO$_3$[20].  Here,
four successive phase transitions are observed as temperature is lowered, from
cubic to
tetragonal to orthorhombic to rhombohedral.  The rhombohedral phase can be
identified as a static JT phase, in which the Ti are all distorted along one of
the eight octahedral directions.  The higher T phases correspond to dynamic JT
phases, in which the Ti tunnels between successively two, four, or all eight
octahedral sites.  In particular, the `cubic' phase is not microscopically of
cubic symmetry, but is a disordered phase with microscopic rhombohedral and
macroscopic average cubic symmetry.  A similar model can apply to LSCO, with a
disordered or dynamic JT phase consisting of a random tilting of the oxygen
octahedra about each Cu, with tunneling between all four allowed tilts in the
HTT phase, between two in the LTO phase, and with only one tilt (static JT) in
the LTT phase.  Section 7 shows that such a sequence of phase transitions
naturally follows from the Hamiltonian introduced[11] to describe this system.
\par
Recent calculations[11,21] have confirmed the close relation between the
ferroelectric perovskites and the superconducting cuprates.  In [11] it was
pointed out that electron-phonon interaction can lead to a structural
instability (negative harmonic phonon frequency) even in a filled band,
as long as the Fermi level fell between the bonding and antibonding bands
associated with hybridizing atoms.  This is clearly the case in La$_2$CuO$_4$,
and Cohen[21] has shown that Ti-O hybridization is essential for
ferroelectricity in BaTiO$_3$ and PbTiO$_3$.
\bigskip
\noindent{\bf 3. Theory of the Structural Transition}
\smallskip\par
The present calculations are based on the nonlinear electron-phonon Hamiltonian
introduced in Ref. [11].  For convenience,
when referring to equations from References [10] or [11], I will prefix them
with a letter $A$ or $B$ respectively (i.e., Eq. 10 of Ref. [11] will be
referred to as Eq. B10, etc.).
\par
It is convenient to rewrite the Hamiltonian of Ref. [11] as
$$H=H_{str}+H_{ph}+H_e+H_{ps}+H_{es}+H_{ep}.\eqno(1)$$
The various terms of Eq. 1 refer to
\par{$\bullet$} {\bf static strains and acoustic phonons}
$$H_{str}={1\over 2M}\sum_{\lambda q}P(\lambda ,q)P(\lambda ,-q)
+\sum_q\Bigl(C_+e_+(-q)e_+(q)+C_-e_-(-q)e_-(q)+
2C_{66}e_{12}(-q)e_{12}(q)\Bigr).\eqno(2)$$
Here, $e_{ij}$ is the usual strain tensor, the elastic constants are
$C_{ij}$[6], and $e_{\pm}=(e_{11}\pm e_{22})/2$, $C_{\pm}=C_{11}\pm C_{22}$.
The strain tensor includes both static and dynamic parts (both strains and
acoustic phonons), and the $P(\lambda ,q)$'s are the momenta associated with
the acoustic phonons.  In the HTT phase, there may be static strains due
to (anisotropic) thermal expansion: $e_{11}\ne 0$.  A non-vanishing $e_{12}$
arises in the LTO phase.  In the HTT and LTO phases, $e_{11}=e_{22}$.  This is
also true in the LTT phase of LBCO, due to interlayer strains[11], but need not
be true in general.  To simplify the present calculations, I will apply them to
a `single layer' model of LBCO, for which $e_-\ne 0$ in the LTT phase.
\bigskip
\par{$\bullet$} {\bf electronic modes}
\smallskip\par
In the vHs model, there is a single, hybridized Cu-O band at the Fermi level.
However, as discussed in Section 2, it is convenient to treat the two vHs as
independent, since they have opposite responses to $e_-$ strains.  Hence,
the carriers will be assumed to belong to group 1 or 2,
depending on whether they are nearer the X- or Y-point vHs, respectively.
In Appendix I, this assumption will be justified by calculating such subbands
from a renormalized microscopic Hamiltonian of the hybridized Cu-O planes.
The electronic energy may be written as
$$H_e=\sum_{i,k} E_i(k)a_{ik}^{\dagger}a_{ik},\eqno(3)$$
with i=1,2.  It is convenient to define the integrated density of states for
each vHs as
$$\rho_{ij}(q)=\sum_k<a_{ik}^{\dagger}a_{j,k+q}>,\eqno(4a)$$
$$\rho_{\pm}(q)=\rho_{11}(q)\pm\rho_{22}(q).\eqno(4b)$$
In the HTT and LTO phases, $\rho_{11}=\rho_{22}$, whereas they are unequal (the
JT splitting) in the LTT phase.  The term involving $\rho_{12}$, which is in
general allowed by symmetry, corresponds to inter-vHs coupling.  This term is
related to umklapp scattering between the two vHs, and can act as a competing
mechanism to the dynamic JT effect.  In Section 5 it will be shown that such a
term could drive a structural transition within a purely orthorhombic phase.
However, it will further be shown that $\rho_{12}$ is symmetry forbidden at the
vHs in the uniform LTO phase, in the absence of spin-orbit coupling.
\par
In the present paper, only intraband inter-vHs coupling will be considered,
with $\rho_{12}$ coupling the vHs's separated by wave vector $Q_0=(\pi
/a,\pi /a)$, as in Refs. [10,11].  It should, however, be noted that similar
considerations could be applied to interlayer coupling, as discussed in Section
2 (Fig. 1b).
\bigskip
\par{$\bullet$} {\bf optical (tilting) modes}
\smallskip\par
Tilting of the oxygen octahedra can be described in terms of (pseudo)rotation
operators, for rotations about the in-plane x- and y-axes[11,22], with $R_x$
and
$R_y$ being the magnitude of the tilt.  To simplify the resulting expressions,
it is convenient to define some auxilliary pair tilting operators.  Thus,
$$R_{ij}(q,q^{\prime})=R_i(q+q^{\prime})R_j(q-q^{\prime}),\eqno(5a)$$
$${\cal R}_{ij}^{(1)}(q)=\sum_{q^{\prime}}R_{ij}(q,q^{\prime}),\eqno(5b)$$
$${\cal R}_{ii}^{(2)}(q)=\sum_{q^{\prime},j}R_{ii}(q,q^{\prime})
\bar\delta_{ij}(cos(q_ja)-cos(q_j^{\prime}a)),\eqno(5c)$$
with $\bar\delta_{ij}\equiv 1-\delta_{ij}=1$ when $i\ne j$, $i,j = 1,2$.
In this case, $H_{ph}$ becomes[11]
$$H_{ph}={1\over 2}\sum_{q}\Bigl(\sum_{\mu}\bigl(K_{\mu\mu}(q)P_{\mu}(-q)
P_{\mu}(q)\bigr)+\tilde\omega_0^2(q)
(R_{11}(0,q)+R_{22}(0,q))$$
$$+{\Gamma_0\over 4}
[{\cal R}_{11}^{(2)}(q){\cal R}_{11}^{(2)}(-q)+{\cal R}_{22}^{(2)}(q)
{\cal R}_{22}^{(2)}(-q)]$$
$$+\Gamma_0^a[{\cal R}_{11}^{(1)}(q){\cal R}_{11}^{(1)}(-q)
+{\cal R}_{22}^{(1)}(q){\cal R}_{22}^{(1)}(-q)]
+\Gamma_2{\cal R}_{11}^{(1)}(q){\cal R}_{22}^{(1)}(-q)
\Bigr),\eqno(5d)$$
where $P_{\mu}$ is the momentum conjugate to $R_{\mu}$, $\mu$ runs over
1 and 2 ($x$ and $y$), and $K_{\mu\mu}$ is defined in Eq. B17a.
\par{$\bullet$} {\bf electron-strain coupling}
The strain-electronic interaction is
$$H_{es}=\sum_q\Bigl(G_+e_+(-q)\rho_+(q)+G_-e_-(-q)\rho_-(q)
+2G_{66}e_{12}(-q)\rho_{12}(q)\Bigr).\eqno(6)$$
with $G_{\pm}=G_{11}\pm G_{22}$.
The term in $G_{66}$ couples the orthorhombic shear strain $e_{12}$ of the
LTO phase to the umklapp term, $<\rho_{12}>$.
\par{$\bullet$} {\bf strain-tilt coupling}
\par
The optical phonon coupling with strain ($H_{ps}$) can be written
$$H_{ps}=\sum_q\Bigl(F_+e_+(-q){\cal R}_+^{(3)}({q\over 2})
+F_-e_-(-q){\cal R}_-^{(3)}({q\over 2})
+F_{66}e_{12}(-q){\cal R}_0^{(3)}({q\over 2})\Bigr).\eqno(7a)$$
Here, I have introduced the combination
$${\cal R}_{\pm}^{(3)}(q)=cos(q_ya){\cal R}_{11}^{(2)}(q)\pm
cos(q_xa){\cal R}_{22}^{(2)}(q),\eqno(7b)$$
and
$${\cal R}_0^{(3)}(q)=\sum_{q^{\prime}}R_{12}(q,q^{\prime})
(1-cos((q_y+q^{\prime}_y)a))(1-cos(q_x-q^{\prime}_x)a)).\eqno(7c)$$
These equations reduce to the corresponding expressions in Ref. [11] when
$q=0$.  In that reference, the Hamiltonian was evaluated only for wave numbers
near the soft mode, $q=0$, $q^{\prime}$ at one of the vHs (either $(0,\pi /a)$
or $(\pi /a,0)$).
\par{$\bullet$} {\bf electron-tilt coupling}
The optical phonon coupling with electrons ($H_{ep}$) can be written
$$H_{ep}=\sum_q\Bigl(\tilde\alpha_+^e\rho_+(-q){\cal R}_+^{(3)}({q\over 2})
+\tilde\alpha_-^e\rho_-(-q){\cal R}_-^{(3)}({q\over 2})
+\tilde\gamma^e\rho_{12}(-q){\cal R}_0^{(3)}({q\over 2})\Bigr).\eqno(8)$$
In the notation of Ref. [11], $\tilde\alpha_{\pm}^e=(\delta^e\pm\alpha^e)/2m$,
$\tilde\gamma^e=\gamma^e/2m$.  This is a nonlinear electron-phonon coupling,
similar to terms which have recently been introduced[23].
\par
All of the above terms involve a sum over $q$.  In an RPA approximation,
the various $q$'s are decoupled.  The soft mode is associated with $q=0$,
so in many applications the non-0 $q$-terms may be neglected (e.g., only
terms in ${\cal R}_{\pm}^{(3)}(0)$ remain).
Moreover, except for the tilt term, Eq. 5d,
the Hamiltonian separates into three parts, involving ($e_+$,
$\rho_+$, ${\cal R}_+^{(3)}$), ($e_-$, $\rho_-$, ${\cal R}_-^{(3)}$), and ($e_
{12}$, $\rho_{12}$, ${\cal R}_0^{(3)}$).
In Appendix II, a toy model is introduced which allows a similar separation of
the tilt terms, Eq. 5d, as well.
\par
The three groups of terms play very different roles in the structural
transitions, particularly in the presence of dynamic JT effects.  At high
temperatures, corresponding to the HTT phase, there are local tilts of the
octahedra, but with no long-range correlations.  In this case, $<R_x^2+R_y^2>
\ne 0$, while $<R_x>=<R_y>=0$.  Hence, the first set of terms controls short
range order, and is non-vanishing in the HTT phase.  These terms will have a
weak temperature dependence at lower temperatures, which can be neglected to
simplify the study of the dynamics -- i.e., the first group of terms
contributes
a constant value to the Hamiltonian and can be neglected.
\par
The second group of terms is nonzero in the LTT phase, and the last in the
LTO phase.  Hence, these terms describe the dynamic competition between the
LTT and LTO phases.
\bigskip
\noindent{\bf 4. Eliminating the Strain and Acoustic Phonon Modes}
\smallskip\par
The elastic strain tensor $e_{ij}$ has static components associated with static
strains and time-dependent components associated with acoustic phonons[11,22].
However, in studying structural phase transitions, it is often convenient to
treat the strains separately from the acoustic phonons.  For instance,
condensation of an optical mode may induce a static strain.  Consider a line of
corner-shared octahedra: if the interatomic distances remain fixed, a static,
antiferrodistortive tilt distortion of the octahedra will reduce the overall
length of the chain.  This seems to be the case in the cuprates: the octahedral
strain in the LTO phase is found to be a secondary order parameter,
proportional
to the square of the octahedral tilt angle[24].  These strains may be formally
decoupled from the problem, following standard practice[12-14], by defining
$$\tilde e_{p}=e_{p}+\lambda_{p},\eqno(9)$$
($p=-,12$)
and chosing $\lambda_{p}$ in such a way as to eliminate the cross terms
between $e_{p}$ and $\rho_{p},$ $R_iR_j$.  The procedure is carried out in
detail in Appendix II, but for illustrative purposes, a simpler calculation is
given here.  Consider the sub-Hamiltonian
$$H^{\prime}=2C_{66}e_{12}^2+2G_{66}e_{12}\rho_{12}.\eqno(10a)$$
Taking $\lambda_{12}=G_{66}\rho_{12}/2C_{66}$ transforms the Hamiltonian to
$$H^{\prime}=2C_{66}\tilde{e}_{12}^2-{J_{12}\over 2}\rho_{12}^2,\eqno(10b)$$
with $J_{12}=G_{66}^2/C_{66}$.  Thus, the phonon motion is formally
decoupled from the electron, leading to an attractive interaction between
electrons.  However, an electronic phase transition is
accompanied by a static distortion: since when $<\tilde e_{12}>=0$, then
$$<e_{12}>=-{G_{66}\over 2C_{66}}<\rho_{12}>.\eqno(11)$$
The separation of charge and phonon variables is not complete ($\tilde e_{12}$
does not commute with $\rho_{12}$), but it has been argued that the additional
complications of noncommutivity are unimportant (see discussion in Ref. [14],
pp. 24-25, and references cited therein).
The resulting attractive interaction between electrons is very
similar to that found in the density wave calculation[10,25], Eq. 2.8 of Ref.
[25], which, in the present notation, becomes
$$V_{kq}={G_{66}^2\hbar\omega_q/2\over(E_k-E_{k+q})^2-(\hbar\omega_q)^2}
\rightarrow -{G_{66}^2\over 2\hbar\omega_q},\eqno(12)$$
Equations 11 and 12 differ only in the denominator, with one equation
containing
$C_{66}$, the other $\hbar\omega_q$.  This substitution arises quite naturally.
In Eq. 2, the term $e_{12}(q)$ includes both static strains and acoustic
phonons.  If the phonons only were included, the term $2C_{66}e_{12}(-q)e_{12}
(q)\rightarrow c_q^{\dagger}c_q(\hbar\omega_q+1/2)$, where the $c_q$ are phonon
operators.  Elimination of the electron-phonon coupling as above would then
lead to an effective electron-electron coupling, Eq. 11 with $C_{66}\rightarrow
\hbar\omega_q$.
\par
In Appendix II, a toy model Hamiltonian is introduced, which allows the optical
phonon coupling to be eliminated in the
same fashion, leaving a purely electronic Hamiltonian with an attractive
effective electron-electron interaction.  There is also the decoupled
phonon Hamiltonian, which splits into two parts: (1) an
acoustic phonon part which, being purely harmonic, can be neglected; (2) an
optical phonon part, which remains anharmonic, but with renormalized
coefficients.
\bigskip
\noindent{\bf 5. Uniform LTO Phase}
\smallskip\par
\par
Before analyzing the dynamic JT phase, it is important to reexamine the
question of whether a structural transition from the HTT phase to a uniform
phase of microscopic LTO symmetry could somehow be driven by the vHs.
At first sight, it would appear that this question has already been answered in
the negative by the work of Pouget, et al.[9].  However, their analysis did
not consider all possible mechanisms for driving a structural instability.
\par
Electron-phonon interaction can lead to a structural instability if the
structural distortion drives a significant density of electronic states below
the Fermi level, as in the one-dimensional CDW problem, due to Fermi surface
nesting.  In the presence of two degenerate vHs, there are two different ways
in which a large dos could be shifted below the Fermi level.  First, the
structural distortion could split the degeneracy of the two vHs, so that one
vHs is shifted below the Fermi level, the other above.  This is clearly what
happens in the LTT phase of LBCO, Fig. 1a, and is the
basis for the dynamic JT model of the LTO phase, discussed in Sections 6 and 7.
As shown by Pouget, et al.[9], the vHs degeneracy is {\it not} lifted in the
uniform LTO phase.
\par
However, there is an alternate mechanism for structural distortion, which could
in principle provide a static model for a {\it uniform} LTO phase.  In
this mechanism, umklapp scattering couples electronic states on degenerate
bands of the Fermi surface, Fig. 1b, leading to coherent
superpositions of the two states with a corresponding gap between the
superposed
states.  In the LTO phase, there is no splitting of the vHs (Fig. 1b), but the
two vHs are now at symmetry-equivalent points of the Brillouin zone, thereby
allowing inter-vHs umklapp scattering.  By introducing a gap at the vHs, this
mechanism can again stabilize a lattice instability.  In the Equations, 1-8,
this inter-vHs coherence corresponds to a finite expectation value for $\rho_{
12}$, and the umklapp scattering to the $G_{66}$ and $\tilde\gamma^e$ terms.
For the calculations of this section, it will be assumed that the two vHs's are
separated by $Q_0=(\pi /a,\pi /a)$, although, as discussed below Eq. 4, other
$Q$-vectors are possible.
\par
The present section is arranged as follows.  In Section 5a, the umklapp
mechanism is discussed, and it is shown that it is likely to be weak in doped
LSCO: it is symmetry-forbidden in the absence of spin-orbit interaction.
Section 5b will further demonstrate
that the static model cannot explain the sequence of transitions from HTT to
LTO
to LTT in terms of competition between $\rho_{12}$ and $\rho_-$.  There is
indeed competition, but the present calculations suggest an either/or
situation:
if there is a {\it static} structural instability, it will be either to an LTO
phase or to the LTT phase, depending on the relative parameter values.  Thus,
the experimentally observed sequence of phases suggests rather a {\it dynamic}
LTO phase -- as in the perovskite ferroelectrics.  The calculations of Sections
6 and 7 confirm this possibility.  The correct sequence of phase transitions is
found, {\it even though umklapp scattering is explicitly neglected} ($G_{66}$
and $\tilde\gamma^e$ set equal to zero).
\bigskip
\noindent{\bf 5a. Interpretation of $\rho_{12}$}
\smallskip\par
The significance of the term $\rho_{12}$, Eq. 4,
can be clarified by recalling the usual manner in which gaps
in the electronic spectrum open at a Brillouin zone boundary[26].  Umklapp
scattering mixes states at $k$ and $k+Q$.  Thus, the electron at the zone
boundary $k=Q/2$ mixes with that at $k=-Q/2$,
$$\psi =a_+c^{\dagger}_{Q/2}+a_-c^{\dagger}_{-Q/2},\eqno(13a)$$
with the $a$'s determined by the eigenvalue equation
$$(E_{Q/2}-E)a_++U_Qa_-=0,\eqno(13b)$$
$$U_Qa_++(E_{-Q/2}-E)a_-=0,\eqno(13c)$$
where $U_Q$ is a measure of the coupling.  Now in the LTO phase, the X and Y
point vHs's are separated by a reciprocal lattice vector, so a similar coupling
can arise.  In the present formalism, this coupling arises by rederiving Eq.
B24
in the presence of the term in $e_{12}\rho_{12}$.  Neglecting fluctuating
quantities, Eq. B24 becomes
$$i{\partial\over\partial t}a_{nk}=\tilde E_n(k)a_{nk}+\tilde E^*a_{mk}
\bar\delta_{mn},\eqno(14)$$
$$\tilde E_1(k)=E_1(k)+G_-e_-+
\tilde\alpha^e_-R_-^2,\eqno(15a)$$
$$\tilde E_2(k)=E_2(k)-G_-e_-
-\tilde\alpha^e_-R_-^2,\eqno(15b)$$
$$\tilde E^*(k)=2G_{66}e_{12}+\tilde\gamma^eR_1R_2.\eqno(15c)$$
\par
Equation 14 shows that, because of the $\rho_{12}$-term, carriers from the
two vHs interact with each other, in the presence of an orthorhombic strain
($e_{12}\ne 0$ or $R_1=\pm R_2\ne 0$).  In this case, the correct eigenstates
can be found by the Bogoliubov construction.  Defining
$$a_{1k}=cos(\theta )c_{1k}+sin(\theta )c_{2k},\eqno(16a)$$
$$a_{2k}=-sin(\theta )c_{1k}+cos(\theta )c_{2k},\eqno(16b)$$
then the equations of motion for the $c_{ik}$'s are decoupled if
$$tan(2\theta )={2\tilde E^*(k)\over \tilde E_2(k)-\tilde E_1(k)},\eqno(17a)$$
and the eigenenergies become
$$E_{\pm}^*=\tilde E_+\pm\sqrt{\tilde E_-^2+\tilde E^{*2}},\eqno(17b)$$
with $\tilde E_{\pm}=(\tilde E_1\pm\tilde E_2)/2$.  The term $\tilde E_+$ may
be neglected in Eq. 17b, since any term which shifts both electronic bands
equally will be compensated by a corresponding shift of the Fermi level.
Thus, when a static orthorhombic strain appears ($<e_{12}>\ne 0$), a
gap $2\tilde E_-$ arises in the electron spectrum, driving the high
density of states associated with the vHs below the Fermi level.
\par
In fact, however, the terms in $\rho_{12}$ must vanish identically in a uniform
LTO phase, in the absence of spin-orbit interaction.  This follows from the
symmetry group $Bmab$ of the LTO phase, and can be considered as a
generalization of Pouget, et al.'s result[9].  Because the lattice contains
glide planes, the eigenfunctions must be two-fold degenerate on one
face of the Brillouin zone, so no gap ($\rho_{12}\ne 0$) can arise.  However,
spin-orbit interaction splits most of the residual degeneracy, allowing a gap
to open at the vHs.  The group theoretical arguments are discussed in more
detail in Appendix III, and the effect of such spin-orbit coupling on the vHs
will be discussed in Section 8.
\bigskip
\noindent{\bf 5b. Mean Field Transition in Electronic Hamiltonian}
\smallskip\par
The above formalism can also be used to study the competition between static
distortions of LTO vs LTT symmetry.  This is most clearly seen by
analyzing the effective electron-electron coupling terms in Eq. 10b.  To
simplify this analysis, it is convenient to temporarily neglect the tilt
coupling and study just the electronic Hamiltonian.  Alternately, the toy
Hamiltonian of Appendix II can be used to formally eliminate the tilt-electron
coupling.  From Appendix II the effective electronic Hamiltonian is found to be
$$H_{\rho}=-{1\over 2}\sum_q\Bigl(J_-\rho_-(-q)\rho_-(q)
+J_0\rho_{12}(-q)\rho_{12}(q)\Bigr),\eqno(18)$$
with the coupling constants, $J_i$, defined in Eq. II6.  A term in $\rho_+$ has
been neglected in Eq. 18.  From charge conservation, $\rho_+(0)$ must be a
fixed
constant, which can be set equal to zero by adjusting the Fermi level.  Due to
the logarithmic divergence of the electronic susceptibility at the vHs, the
dominant singularity (soft mode) corresponds to $q=0$, and at the RPA level of
approximation, this is the only mode which need be discussed.
\par
The mean field solution can be found easily, as in Section 5a and
Ref. [10].  The mean field Hamiltonian becomes
$$H^e_{MF}=\sum_k\Bigl(E_k(a_{1k}^{\dagger}a_{1k}+a_{2k}^{\dagger}a_{2k})-
D_-(a_{1k}^{\dagger}a_{1k}-a_{2k}^{\dagger}a_{2k})-D_0(a_{1k}^{\dagger}a_{2k}
+a_{2k}^{\dagger}a_{1k})\Bigr),\eqno(19a)$$
with
$$D_-={J_-\over 2}<\rho_-(0)>,\eqno(19b)$$
$$D_0={J_0\over 2}<\rho_{12}(0)>.\eqno(19c)$$
$H^e_{MF}$ can be diagonalized as in Eqs. 16, 17, yielding eigenvalues
$$E_{k\pm}^*=E_k\pm D,\eqno(20)$$
with $D^2=D_-^2+D_0^2$.  The self-consistency conditions, Eqs. 19b,c, yield
equations for the two gaps
$$D_-={J_-D_-\over 2D}\sum_k\bigl(f(E_{k+})-f(E_{k-})\bigr),\eqno(21a)$$
$$D_0={J_0D_0\over 2D}\sum_k\bigl(f(E_{k+})-f(E_{k-})\bigr),\eqno(21b)$$
where $f(E)=1/(exp((E-E_F)/k_BT)+1)$.
Except in the special case $J_-=J_0$, Eqs. 21a,b cannot simultaneously have
nonvanishing solutions.  Instead, the solution with the larger value of $J_i$
prevails.  This makes good physical sense: the two transitions, LTO and LTT,
are both driven by the same dos peak.  Whichever phase is stabilized first
uses up the available dos, and prevents the other from occuring.
\par
In this case, the calculation may be simplified by ignoring terms of the
non-condensing symmetry.  This was in effect what was done in Refs. [10] and
[11].  In Ref. [10], only the LTO solution appeared, while the calculations of
Ref. [11]
neglected terms involving $\rho_{12}$, and predominantly described the LTT
phase.  It should be particularly noted that the present calculation has
reproduced the BCS-like calculations of Ref. [10], but starting from the more
microscopic Hamiltonian of Ref. [11], thereby explicitly displaying the close
connection between the two works.
\par
To compare with the results of Ref. [10], assume the LTO phase is favored,
i.e.,
$J_0>J_-$.  Then $D_-=0$, and the gap $D=D_0$ is given by the solution of
$$1={J_0\over 2D}\int dE N(E)\bigl(f(E-D)-f(E+D)\bigr).\eqno(22)$$
For a logarithmic dos,
$$N(E)={1\over B}ln\bigl({B\over 2E}\bigr),\eqno(23)$$
the zero temperature gap is
$$D(0)={eB\over 2}e^{-B/J_0}.\eqno(24)$$
This should be compared to Eq. 14 of Ref. [26].
Figure 2 illustrates the temperature dependence of the gap, found by solving
Eq.
22-23 numerically.
Note that, based on Eq. 11, there will be a nonvanishing orthorhombic strain in
the LTO phase, $<e_{12}>\propto -<\rho_{12}>$.  Indeed, the present solution
is similar to that found in Ref. [11], Model 2, except that inclusion of
umklapp
processes drives the large vHs dos below the Fermi level.
\par
This section has explored the role of the term $\rho_{12}$ in stabilizing a
static LTO phase.  However, symmetry arguments suggest that such terms are
small in the absence of spin-orbit coupling.  Hence, in Sections 6 and 7, the
opposite limit will be explored.  Terms in $\rho_{12}$ will formally be
retained in the Hamiltonian, but I will attempt to determine under what
circumstances an LTO-type phase might arise when the $\rho_{12}$ terms are
small
or vanishing.  It will be shown that the LTO phase can be interpreted as a
dynamic JT phase.
\bigskip
\noindent{\bf 6. Dynamic JT Hamiltonian}
\smallskip\par
\bigskip
\noindent{\bf 6a. Real-Space Hamiltonian: Intracell and Intercell Coupling}
\smallskip\par
This section will present a more accurate treatment of the combined
electron-optical phonon Hamiltonian, following a conventional treatment of the
dynamic JT effect. It is convenient to first transform the Hamiltonian, Eq.
1, back into real space:
$$H=\sum_{l,\eta}\Bigl({\Gamma_0\over 4}[({R_1({\vec l})-R_1({\vec l}+
\eta a\hat y)\over 2})^4+({R_2({\vec l})-R_2({\vec l}+\eta a\hat x)\over 2})^4]
+{\Gamma_0^a\over 2}[R_1^4({\vec l})+R_2^4({\vec l})]
+{\Gamma_2\over 2}R_1^2({\vec l})R_2^2({\vec l})$$
$$+(\tilde\alpha_+^{e}\rho_+({\vec l})+F_+e_+({\vec l}))
[({R_1({\vec l})-R_1({\vec l}+\eta
a\hat y)\over 2})^2+({R_2({\vec l})-R_2({\vec l}+\eta a\hat x)\over 2})^2]$$
$$+(\tilde\alpha_-^{e}\rho_-({\vec l})+F_-e_-({\vec l}))
[({R_1({\vec l})-R_1({\vec l}+\eta
a\hat y)\over 2})^2-({R_2({\vec l})-R_2({\vec l}+\eta a\hat x)\over 2})^2]$$
$$+({\tilde\gamma^{e}\rho_{12}({\vec l})+F_{66}e_{12}({\vec l})
\over 4})\sum_{\eta^{\prime}}
[R_1({\vec l})-R_1({\vec l}+\eta\hat ya)][R_2({\vec l})
-R_2({\vec l}+\eta^{\prime}\hat xa)]$$
$$+{\tilde\omega_0^2\over 2}(R_1(\vec l)^2+R_2(\vec l)^2)
+C_+e_+^2(\vec l)+C_-e_-^2(\vec l)+2C_{66}e_{12}^2(\vec l)$$
$$+G_+e_+(\vec l)\rho_+(\vec l)+G_-e_-(\vec l)\rho_-(\vec l)
+2G_{66}e_{12}(\vec l)\rho_{12}(\vec l)\Bigr),\eqno(25)$$
where $\eta$ and $\eta^{\prime}$ are summed over $\pm 1$ and
$$\rho_-({\vec l})=a_{1l}^{\dagger}a_{1l}-a_{2l}^{\dagger}a_{2l},\eqno(26)$$
with a similar expression for $\rho_{12}$.  To form the full Hamiltonian, Eq.
1, the electronic energy, Eq. 3, and the kinetic energy terms of Eqs. 2 and 5d
must be added to Eq. 25.  Moreover, a possible $q$-dependence of $\tilde\omega_
0$ has been neglected.  Eliminating the strain terms from Eq. 25, as in
Appendix
II, yields
$$H=\tilde H_{str}+\sum_{l,\eta}\Bigl({\Gamma_0^{\prime}\over 4}
[({R_1({\vec l})-R_1({\vec l}+
\eta a\hat y)\over 2})^4+({R_2({\vec l})-R_2({\vec l}+\eta a\hat x)\over 2})^4]
+{\Gamma_0^a\over 2}[R_1^4({\vec l})+R_2^4({\vec l})]
+{\Gamma_2\over 2}R_1^2({\vec l})R_2^2({\vec l})$$
$$+\tilde\alpha_+^{e\prime}\rho_+({\vec l})[({R_1({\vec l})-R_1({\vec l}+\eta
a\hat y)\over 2})^2+({R_2({\vec l})-R_2({\vec l}+\eta a\hat x)\over 2})^2]$$
$$+\tilde\alpha_-^{e\prime}\rho_-({\vec l})[({R_1({\vec l})-R_1({\vec l}+\eta
a\hat y)\over 2})^2-({R_2({\vec l})-R_2({\vec l}+\eta a\hat x)\over 2})^2]$$
$$+{\tilde\gamma^{e\prime}\over 4}\rho_{12}({\vec l})\sum_{\eta^{\prime}}
[R_1({\vec l})-R_1({\vec l}+\eta\hat ya)][R_2({\vec l})
-R_2({\vec l}+\eta^{\prime}\hat xa)]$$
$$+{\tilde\omega_0^2\over 2}(R_1(\vec l)^2+R_2(\vec l)^2)
-{J_+^{\prime}\over 2}\rho_+^2({\vec l})
-{J_-^{\prime}\over 2}\rho_-^2({\vec l})
-{J_0^{\prime}\over 2}\rho_{12}^2({\vec l})\Bigr)+H^*
,\eqno(27a)$$
where $\Gamma_{0}^{\prime}=\Gamma_0-F_+^2/C_+-F_-^2/C_-$, $\tilde\alpha_{\pm}
^{e\prime}=\tilde\alpha_{\pm}^e-G_{\pm}
F_{\pm}/4C_{\pm}$,
$\tilde\gamma^{e\prime}=\tilde\gamma^e-G_{66}F_{66}/2C_{66}$,
$J_{\pm}^{\prime}=G_{\pm}^2/2C_{\pm}$, and $J_0^{\prime}=G_{66}^2/C_{66}$;
$\tilde H_{str}$ is a quadratic pseudostrain Hamiltonian (Eq. IIa1) decoupled
from the remaining terms and of no further interest; and
$$H^*=-{1\over 32}\sum_l\Bigl(\Gamma_{0+}
[({R_1({\vec l})-R_1({\vec l}+a\hat y)})^2
({R_1({\vec l})-R_1({\vec l}-a\hat y)})^2+$$
$$({R_2({\vec l})-R_2({\vec l}+a\hat x)})^2
({R_2({\vec l})-R_2({\vec l}-a\hat x)})^2]$$
$$+\Gamma_{0-}
[({R_1({\vec l})-R_1({\vec l}+a\hat y)})^2+
({R_1({\vec l})-R_1({\vec l}-a\hat y)})^2]
[({R_2({\vec l})-R_2({\vec l}+a\hat x)})^2+
({R_2({\vec l})-R_2({\vec l}-a\hat x)})^2]$$
$$+\Gamma_{00}\bigl(\sum_{\eta ,\eta^{\prime}}
[R_1({\vec l})-R_1({\vec l}+\eta\hat ya)][R_2({\vec l})
-R_2({\vec l}+\eta^{\prime}\hat xa)]\bigr)^2
\Bigr),\eqno(27b)$$
with $\Gamma_{0\pm}=F_+^2/C_+{\pm}F_-^2/C_-$, $\Gamma_{00}=F_{66}^2/8C_{66}$.
\par
The role of the strain forces can be determined by comparing Eqs. 25 and 27.
In addition to the effective electron-electron interaction terms, the
strain has introduced longer-range tilt-tilt interactions.  All of the terms of
Eq. 25 are either on-site or nearest neighbor interactions, except the terms in
$\tilde\gamma^e$ and $F_{66}$.  In contrast, all of the terms of $H^*$ involve
further neighbor interactions.
\par
In a mean field treatment, it is assumed that there is a nonzero tilt present
on
each lattice site, even in the HTT phase
$$<R_1^2({\vec l})+R_2^2({\vec l})> =\bar R^2\ne 0,\eqno(28a)$$
independent of $\vec l$.  Then the dynamic variable is the tilt direction,
$\phi_l$, with
$$R_1({\vec l})=(-1)^{i+j}\bar Rcos\phi_l,\eqno(28b)$$
$$R_2({\vec l})=(-1)^{i+j}\bar Rsin\phi_l,\eqno(28c)$$
with ${\vec l}=(ia,ja)$.  Note the factor $(-1)^{i+j}$ -- this is introduced
because the intercell coupling must locally be antiferrodistortive, due to the
corner sharing of the octahedra.  By explicitly taking out this factor, it can
be expected that $\phi_l$ will be a smooth function of position, with a
well behaved continuum limit.
\par
Equation 27 can be separated into intracell and intercell terms, as $H=\sum_l
(H^o_0+H^o_1+H^o_2)$, with
$$H^o_0={1\over 2}(\Gamma_0^{\prime\prime}+\Gamma_2^{\prime\prime})\bar R^4
+({\tilde\omega_0^2\over 2}+{\tilde\alpha_+^{e\prime}\over 2}\rho_+)\bar R^2
-{J_+^{\prime}\over 2}\rho_+^2,\eqno(29a)$$
$$H^o_1=-{1\over 2}\Gamma_2^{\prime\prime}\bar R^4cos4\phi
+{\tilde\alpha_-^{e\prime}\over 2}\rho_-\bar R^2cos2\phi
+{\tilde\gamma^{e\prime}\over 2}\rho_{12}\bar R^2sin2\phi$$
$$-{J_-^{\prime}\over 2}\rho_-^2
-{J_0^{\prime}\over 2}\rho_{12}^2
,\eqno(29b)$$
with $\Gamma_0^{\prime\prime}=\Gamma_0^a+(\Gamma_0^{\prime}-\Gamma_{0+})/16$
and $\Gamma_2^{\prime\prime}=(\Gamma_2-2\Gamma_0^{\prime\prime}-\Gamma_{00}-
\Gamma_{0-}/4)/8$.  Since all terms refer to the same cell, the
$\vec l$-dependence is not explicitly displayed.  Here,
$H^o_0$ is $\phi$-independent, $H^o_1$ is a single cell Hamiltonian, while
$H^o_2$ incorporates the intercell coupling.  The intercell coupling term
is complicated, and will be explicitly displayed only at the mean field
level, for which $cos\phi(\vec l^{\prime})=<cos\phi >$ and $sin\phi(\vec l^
{\prime})=<sin\phi >$ for all $\vec l^{\prime}\ne\vec l$.  In this case,
$H_2^o=
H_2^{\prime}(<cos\phi >,<sin\phi >)
-H_2^{\prime}(0,0)$, with
$$H_2^{\prime}(<cos\phi >,<sin\phi >)=
{\Gamma_0^{\prime}-\Gamma_{0+}\over 32}\bar R^4\bigl[(cos\phi +<cos\phi >)^4
+(sin\phi +<sin\phi >)^4\bigr]$$
$$-{\Gamma_{0-}+4\Gamma_{00}\over 8}\bar R^4(cos\phi +<cos\phi >)^2(sin\phi
+<sin
\phi >)^2$$
$$+{\tilde\alpha_+^{e\prime}\rho_+\bar R^2\over 2}\bigl[(cos\phi +<cos\phi >)^2
+(sin\phi +<sin\phi >)^2\bigr]$$
$$+{\tilde\alpha_-^{e\prime}\rho_-\bar R^2\over 2}\bigl[(cos\phi +<cos\phi >)^2
-(sin\phi +<sin\phi >)^2\bigr]$$
$$+\tilde\gamma^{e\prime}\rho_{12}\bar R^2(cos\phi +<cos\phi >)(sin\phi +<sin
\phi >).\eqno(29c)$$
\par
As discussed in [11], the terms in $\rho_+$ are non-critical, and can be
eliminated from Eq. 27.  Thus, $\rho_+(q=0)$ is just the number of holes in
the conduction band, $\rho_+(0)=1$ in the present case, where $\rho_-(0)$ (or
$\rho_{12}$) becomes non-zero only in the low-T phase, and hence can be taken
as
an order parameter of the transition.  Thus, we may assume $\rho_+(q)\sim\rho_+
(0)$, and eliminate the $\rho_+$-dependent terms from Eq. 27.  The harmonic
phonon frequency is renormalized  $\omega_0^2=\tilde\omega_0^2+
\tilde\alpha_+^{e\prime}(\rho_+(0)-2)$.  In this equation, I have incorporated
an additional correction[11], the term in $-2$, due to the filled, bonding
band of the hybridized Cu-O band.  This term is important in destabilizing the
lattice, $\omega_0^2<0$, both in the present problem and in ferroelectrics and
other structural instabilities.
\par
While the $\rho_+$-terms are neglected in the present analysis, they may yet
have important effects in these materials.  They provide a coupling between the
tilt and the {\it local average electronic density}, and lead to the
possibility
of a {\it microscopically heterogeneous phase}.  Such nanoscopic disorder has
previously been suggested to play an important role in doping these materials
away from the vHs[28], while a related phase heterogeneity has been proposed to
arise on doping away from the antiferromagnetic phase at half-filling[17].
\bigskip
\noindent{\bf 6b. Intracell Hamiltonian: Static JT Effect}
\smallskip\par
It is convenient to begin by discussing the JT effect within a single cell.
The terms in $H^o_0$ lead to a tilting of the octahedron, $\bar R\ne 0$, but
with no preferred orientation.  If $\bar R$ is assumed to be a fixed constant,
then the JT effect involves the angular orientation of the tilted octahedron,
$\phi$.  For the single cell problem, this involves $H^o_1$, Eq. 29b.  The
solution to Eq. 29b depends on whether the system is in the static or dynamic
JT limit.  In the dynamic limit, the tilt kinetic energy operator associated
with $\phi$ must be added to Eq. 29b; in the opposite limit, the tilts are
static, and the kinetic energy can be neglected.  In this static case, the
electronic operators can be diagonalized by a transformation similar to Eq. 16.
Alternatively, a pseudospin formalism[12-14,29] (Appendix IV) can be employed.
In the two-dimensional subspace spanned by the electronic operators $a_1^{
\dagger}(\vec l)$, $a_2^{\dagger}(\vec l)$, the $\rho$ operators can be
represented by Pauli matrices
$$\rho_-=\sigma_z,$$
$$\rho_{12}=\sigma_x;$$
hence, the $\rho^2$-terms in Eq. 29b reduce to constants, and can be
eliminated.
Transforming the electronic states by Eq. 16, the eigenenergies of Eq. 29b
become
$$E_{1\pm}^o=-{1\over 2}\Gamma_2^{\prime\prime}\bar R^4cos4\phi
\pm {1\over 2}\sqrt{\tilde\alpha_-^{e\prime 2}\bar R^4cos^22\phi
+\tilde\gamma^{e\prime 2}\bar R^4sin^22\phi}.\eqno(30)$$
Equation 30 demonstrates the JT effect in the present system.  If $E_{1-}^o$
were minimized with respect to $\bar R$, the electronic term would always give
rise to a non-vanishing JT distortion ($\bar R\ne 0$); incorporation of other
terms from the original Hamiltonian of order $\sim \bar R^4$ would not change
this result.  However, since the electron-phonon coupling is quadratic in $R$,
there are also harmonic terms in the Hamiltonian of order $R^2$, and a JT
splitting will arise only if the coupling coefficient (e.g., $\alpha^{e\prime}
$) is large enough (or the harmonic coefficient is negative -- see [11]).
\par
Equation 30 represents a potential with four degenerate energy minima.
Depending on the parameter values, these minima may lie either along the
$\pm x$ and $\pm y$ axes, or at $45^o$ to this.
It is expected that $\tilde\alpha_-^{e\prime}>\tilde\gamma^{e\prime}$; if
$\tilde\gamma^{e\prime}=0$, the minima would lie along the principal axes when
$-2\Gamma_2^{\prime\prime}\bar R^2-|\tilde\alpha_-^{e\prime}|<0$, and at $45^o$
when the inequality is reversed.  In the static limit, the tilted octahedron
will be located at one of the four equivalent minima.  In the absence of
intercell coupling, there can be no macroscopic phase transition: at low
temperatures the octahedra will be randomly distributed among all four
minima; as the temperature is raised, the octahedra can hop among the various
minima.
\par
In what follows, it will be convenient to approximate the lower JT solution,
$E_{1-}^o$ of Eq. 30 by a simpler form
$$E_{1-}^o\simeq E_a-E_bcos4\phi.\eqno(31)$$
For instance, if $\tilde\gamma^{e\prime}=0$, $2E_b\simeq|\tilde\alpha_-^{e
\prime}|\bar R^2/2+\Gamma_2^{\prime\prime}\bar R^4$.  For nonzero $\tilde\gamma
^{e\prime}$, the angle dependence is a function of $\tilde\alpha_-^{e\prime}-
\tilde\gamma^{e\prime}$. The terms which comprise $E_b$ tend to appear as the
difference between two quantities (e.g., $\Gamma_0^{\prime}$, $\tilde\alpha_-^
{e\prime}-\tilde\gamma^{e\prime}$).  This occurs
because for an isolated free molecule the `JT distortions' amount to a
pure rotation of the molecule; the distortion is a solid-state effect,
due purely to crystalline anisotropy (this fact appears to have been first
pointed out in Ref. [30]).  The constant $E_a$ can be absorbed into the
angle-independent part of the Hamiltonian, $H_0^o$.
\bigskip
\noindent{\bf 6c. Dynamic JT Effect and Relation to Conventional ($E\otimes e$)
JT Effect}
\smallskip\par
When dynamic effects are important, the ionic kinetic energy operator from Eq.
5d must be retained in the Hamiltonian, Eq. 29b.  The exact eigenstates no
longer can be written in Born-Oppenheimer form, but are of the form
$$\Psi_n(\bar R,\phi )=\chi_{n1}(\bar R,\phi )\psi_-(\bar R,\phi )+
\chi_{n2}(\bar R,\phi )\psi_+(\bar R,\phi ),$$
where $\psi_{\pm}$ is the electronic wavefunction corresponding to the
energy $E_{1\pm}^o$, Eq. 30, and the $\chi_{ni}$ are wavefunctions of the
nuclear motions.  In the limit of strong JT coupling, the upper JT level can be
neglected, and a Born-Oppenheimer wave function is approximately recovered:
$$\Psi_n=f(\bar R)\phi_n(\phi )\chi_{\pm}(\bar R,\phi ),$$
where $\chi_{\pm}$ is the lower energy electronic wave function and the
nuclear wave function has been separated into a radial part, $f(\bar R)$,
assumed to be approximately constant, and an angular part, $\phi_n$, which
satisfies
$${-\hbar^2\over 2\bar R^2}\bigl({\partial^2\over\partial\phi_l^2}+\beta
cos(4\phi_l)\bigr)\phi_n=E_n\phi_n,\eqno(32a)$$
where the first term in Eq. 32a is the angular contribution to the tilt kinetic
energy, and
$${\hbar^2\beta\over 2\bar R^2}=E_b.\eqno(32b)$$
This distorted octahedron, strongly
and nonlinearly coupling the tilt and the electronic state, constitutes the
polaron of the present problem.  For convenience, this tilt polaron will be
referred to as a `cageon'.
\par
Equation 32a is a form of Matthieu's equation[31-33] which often arises in the
dynamic JT problem.  The vibronic potential here has four minima, along the
positive and negative x- and y-axes (for $\beta >0$), corresponding to the
four possible static JT tilts of the CuO$_6$ octahedron.  When $\beta$ is very
large, the lowest-energy state is fourfold degenerate, corresponding to these
static distortions, with a weak tunneling
between the states.  For smaller barriers, this quartet of states breaks up
into a pair of doublets.  The solutions of Eq. 32a are discussed further in
Appendix V.
\par
In the closely related $E\otimes e$ JT effect[12-14], a similar Mathieu's
equation arises, but with three-fold degeneracy (corresponding to elongation of
the octahedron along an x, y, or z axis).  This problem is often simplified by
approximating the intercell coupling by a quadratic form.  In this case, the
weak tunneling limit reduces to a three-states Potts model[34,35]. For the
present,
four-fold degenerate model, this would correspond either to a pseudo-spin 3/2
system, or to a four-states Potts model.
\bigskip
\noindent{\bf 7. Intercell Coupling}
\smallskip\par
\bigskip
\noindent{\bf 7a. High-Temperature Limit (Disordered JT Phase)}
\smallskip\par
In the intercell coupling term, $<sin\phi >$ and $<cos\phi >$ are independent
variables, constrained by $(<sin\phi >)^2+(<cos\phi >)^2\le 1$.  Two classes of
solution are of particular interest: $<cos\phi >\ne 0$, $<sin\phi >=0$, the
LTT solution, and $<sin\phi >=<cos\phi >\ne 0$, the LTO solution.
\par
The analysis is most straightforward in the weak tunneling limit, when only
the four lowest-lying levels of each octahedron need be considered.  In the
high-temperature limit, $<sin\phi >=<cos\phi >=0$, and only the single cell
Hamiltonian, Eq. 29a,b, need be considered.  If tunneling is completely
absent, the four wave functions are each localized in one of the potential
minima of $cos4\phi$.  Near the minimum, the potential is quadratic in $\phi$,
and the wave functions are well approximated by harmonic oscillator wave
functions.  In particular, the ground state will be approximately a Gaussian
wave function,
$$\psi =\sqrt{2\nu\over\pi}e^{-\nu\phi^2},\eqno(33)$$
with $\nu=\sqrt{2\beta}$.
Inclusion of overlap between Gaussians centered on different wells splits the
degeneracy.  However, just as in the 3-well model, it is important to
recognize that the total wave function is a combined electron-phonon wave
function.  The electronic wave functions are only symmetric under a $4\pi$
rotation:  $\chi (\phi +2\pi )=-\chi (\phi )$.  Thus, the {\it nuclear} wave
function must also satisfy
$$\phi_n(\phi +2\pi )=-\phi_n(\phi ),\eqno(34)$$
so that the total wave function $\Psi_n$ has $2\pi$ symmetry.  In the n=3
case, this changes the sign of the overlap, thereby reversing the order of the
levels[12,31,36].  This sign change is now recognized to be an example of
Berry's phase[37].
\par
In a perturbation calculation of the four-well problem, this means
diagonalizing
an 8$\times$8 matrix.  However, since the wavefunctions 5-8 are just the
negative of wave functions 1-4, this immediately reduces to a 4$\times$4 matrix
$$\left(\matrix{H_{11}-E&H_{12}-SE&0&-(H_{12}-SE)\cr
         H_{12}-SE&H_{11}-E&H_{12}-SE&0\cr
         0&H_{12}-SE&H_{11}-E&H_{12}-SE\cr
         -(H_{12}-SE)&0&H_{12}-SE&H_{11}-E}\right)=0,\eqno(35)$$
with $S$ the wave function overlap, $H_{12}$ the nearest-neighbor Hamiltonian
overlap, and $H_{11}$ the diagonal Hamiltonian.  For the present problem, these
matrix elements are
$$H_{11}=\alpha\nu (1-{\nu\over 2}e^{-2/\nu}),\eqno(36a)$$
$$H_{12}=S\alpha\nu (1+{\nu\over 2}(e^{-2/\nu}-{\pi^2\over 2})),\eqno(36b)$$
$$S=e^{-\nu\pi^2/8},\eqno(36c)$$
with $\alpha={\hbar^2/2\bar R^2}$.  This overlap splits the four degenerate
levels into two pairs of levels, with energies
$$E_{\pm}={H_{11}\pm\sqrt{2}H_{12}\over 1\pm \sqrt{2}S}.\eqno(36d)$$
The wave functions
associated with $E_+$ are in the subspace spanned by the vectors $(1/\sqrt{2},
1,1/\sqrt{2},0)$ and $(0,1/\sqrt{2},1,1/\sqrt{2})$, where the various elements
refer to the amplitudes in the separate minima, while the $E_-$ wave functions
correspond to $(1/\sqrt{2},-1,1/\sqrt{2},0)$ and $(0,1/\sqrt{2},-1,1/\sqrt{2}
)$.  These latter are higher in energy by a factor
$$\Delta E=E_--E_+={2\sqrt{2}\alpha\nu^2S(\pi^2/4-e^{-2/\nu})\over 1-2S^2}.
\eqno(37)$$
As $S\rightarrow 0$, $\Delta E$ vanishes, leading to a static, but
disordered JT phase, with the octahedra equally likely to have any of the
four tilt distortions.  The resulting lattice symmetry is pseudo-cubic, in that
all orientations are equally likely.  This four well approximation becomes
exact as $\nu\rightarrow\infty$, and Figure 3 shows that it remains
qualitatively correct for all values of $\nu$.  The solid lines in Fig. 3 are
the exact eigenvalues of Eq. 32a, as discussed in Appendix V.  The axes of Fig.
3 are in normalized units, introduced in Appendix V: $\nu =4\sqrt{q}$, $E_\pm =
4\alpha a$.
\par
There appears to be an asymmetry in the above wavefunctions, since they are
all centered on wells 2 or 3.  However, linear combinations of these
wavefunctions can be generated which are centered on the other wells.  For
example, $\sqrt{2}(0,1/\sqrt{2},1,1/\sqrt{2})-(1/\sqrt{2},1,1/\sqrt{2},0)
=(-1/\sqrt{2},0,1/\sqrt{2},1)$.
\bigskip
\noindent{\bf 7b. Phase Transitions: Dynamic JT Phases}
\smallskip\par
As the temperature is lowered, intercell coupling will lead to an ordered
low temperature phase, with $<cos\phi >$ or $<sin\phi >$ non zero.  A complete
solution of Eq. 29 is prohibitively difficult, particularly since most of
the parameters are not well known.  In this section, a number of
simplifications
are introduced to make the problem more tractable, allowing a determination of
the conditions under which an LTO (dynamic JT) phase might be stable.  These
simplifications are: (1) since a key question will be to establish whether an
LTO-like phase can be stabilized in the absence of strong umklapp scattering,
it will be assumed that $\tilde\gamma^{e\prime}=0$.  (2) It will be assumed
that $<cos\phi >$ and $<sin\phi >$ are small near the transition (i.e., that
the transition is second order, or weakly first order), so that $H_2^o$ can be
linearized in these quantities.
\par
(3) Even with these assumptions, $H_2^o$ remains complicated:
$$H_2^o={[\Gamma_0^{\prime}-\Gamma_{0+}+2\Gamma_{0-}+8\Gamma_{00}]
\over 8}\bar R^4(cos^3\phi <cos\phi >+sin^3\phi <sin\phi >)$$
$$+[\tilde\alpha_+^{e\prime}-(\Gamma_{00}+{\Gamma_{0-}
\over 4})\bar R^2]\bar R^2(cos\phi <cos\phi>+sin\phi <sin\phi >)$$
$$+\tilde\alpha_-^{e\prime}\rho_-\bar R^2(cos\phi <cos\phi>-sin\phi <sin\phi >)
.\eqno(38a)$$
In the same spirit of replacing Eq. 30 by Eq. 31, this may be replaced by
the simpler form
$$H_2^o=H_0^{\prime}(cos\phi <cos\phi>+sin\phi <sin\phi >),\eqno(38b)$$
which involves only one unknown parameter, $H_0^{\prime}$.  To see how good
this approximation is, consider the case in which the tilting Hamiltonian, $H_
{tilt}\equiv H_1^o+H_2^o$ of Eq. 29, is dominated by the vHs JT effect (i.e.
the terms involving $\tilde\alpha_-^{e\prime}$).  In this case, the electronic
term can be integrated out as in Eq. 30, leading to
$$H_{tilt}=-{\tilde\alpha_-^{e\prime}\bar R^2\over 2}\bigl|(cos\phi +<cos\phi >
)^2-(sin\phi +<sin\phi >)^2\bigr|.\eqno(39)$$
Figure 4 shows that even for this highly singular potential, the simpler form
of Eq. 31 plus 38b provides a reasonable approximation.  In approximating Eq.
39, the parameter $H_0^{\prime}$ must be chosen to be negative; hence, only
this regime will be explored in detail below.
\par
With the intercell coupling given by Eq. 38b, the perturbation matrix may be
recalculated.  The Hamiltonian matrix, Eq. 35, becomes
$$\left(\matrix{x+AH_1&y+(A-B)H_2&0&-(y+(A+B)H_2)\cr
         y+(B-A)H_2&x+BH_1&y+(A+B)H_2&0\cr
         0&y-(A+B)H_2&x-AH_1&y+(B-A)H_2\cr
         -(y-(A+B)H_2)&0&y+(A-B)H_2&x-BH_1}\right)=0,\eqno(39)$$
where $x=H_{11}-E$, $y=H_{12}-SE$, $A=<cos\phi >$, $B=<sin\phi >$, $H_1=H_0^
{\prime}exp(-1/8\nu )$, and $H_2=H_1S/\sqrt{2}$.  For arbitrary $<sin\phi >\ne
0$, Eq. 39 must be diagonalized numerically.  However, for the LTT phase ($<
sin\phi >=0$), the solutions can be found analytically.  The potential wells
are labelled in such a way that the $<cos\phi >$ terms lower the state $(1,0,
0,0)$ (state $\psi_1$) and raise the state $(0,0,1,0)$ ($\psi_3$) in energy.
Then, in the LTT phase ($B=0$), two solutions have $\psi_1=0$, with energies
$$E_{1,2}={2S\bar H-AH_1/2\pm\sqrt{2\bar H^2+A^2(H_1^2/4-2H_2^2(1-2S^2))-2S\bar
HAH_1}\over 1-2S^2};\eqno(40a)$$
while the other two have $\psi_3=0$ and
$$E_{3,4}={2S\bar H+AH_1/2\pm\sqrt{2\bar H^2+A^2(H_1^2/4-2H_2^2(1-2S^2))+2S\bar
HAH_1}\over 1-2S^2}.\eqno(40b)$$
In the above equations, $E$ is measured from $H_{11}$, and $\bar
H=H_{11}S-H_{12
}$.  Figure 5 illustrates how the energies in the four wells vary as a function
of $<cos\phi >$, for several values of $S$, both for the LTT phase ($<sin\phi
>=
0$ -- dashed lines) and for the LTO phase ($<sin\phi >=<cos\phi >$ -- solid
lines).  [Note that there is a break in slope in the LTT phase at $(<cos\phi >)
^2=1/2$, since the constraint $(<sin\phi >)^2+(<cos\phi >)^2\le 1$ comes into
play.  In the figure, it is assumed that $<sin\phi >=\sqrt{1-(<cos\phi >)^2}$
in
this case.  However, this regime has no immediate relevance, since $<cos\phi >$
is always found to be $<0.7$ in the numerical calculations for the LTO phase.]
\par
{}From Figure 5, the energy is always lowered when $<cos\phi >\ne 0$, and at
low
enough temperatures there will be a transition to a dynamic JT phase.
In mean field theory, the phase diagram may readily be calculated from the
self-consistency condition for $<cos\phi >$:
$$<cos\phi >={Tr cos\phi e^{-H/k_BT}\over Tr e^{-H/k_BT}},\eqno(41)$$
where $Tr$ stands for the trace.  In the four-state model, $cos\phi$ vanishes
on
average in wells 2 and 4, and $cos\phi =\pm e^{-1/8\nu}$ in wells 1(+) and
3(--).  Thus
$$<cos\phi >=e^{-1/8\nu}{\sum_j(p_{1j}-p_{3j})e^{-E_j/k_BT}
\over e^{-E_1/k_BT}+e^{-E_2/k_BT}+e^{-E_3/k_BT}+e^{-E_4/k_BT}},\eqno(42)$$
where $p_{ij}$ is the probability that well $i$ is occupied in the state of
energy $E_j$.  Note that any term common to all the $E_i$'s cancels out of the
ratio in Eq. 42.  Thus, Eq. 42 depends on three parameters, $S$ (or,
equivalently $\nu$), $\bar H$, and $H_0^{\prime}$.  At sufficiently high
temperatures, Eq. 42 has no non-zero solutions.  As T is lowered, there is a
critical temperature below which there are values of $<cos\phi >\ne 0$.  Figure
6 illustrates the solution for the LTT phase, plotting the right-hand side of
Eq. 42 against $<cos\phi >$ at a number of different temperatures.  The
solutions of Eq. 42 correspond to the intersection of the solid and dashed
lines.  These solutions lead to the evolution of the ordered phases, $<cos\phi
>(T)$ shown in Figure 7.
\bigskip
\noindent{\bf 7c. Phase Transitions: LTO vs LTT Phase}
\smallskip\par
Figure 7 compares the resulting phase diagrams for both the LTT and the LTO
phases, for several sets of parameters.  The parameters were chosen
to approximately match the critical temperature of the LTO phase in La$_2$CuO$
_4$.  Figure 8 shows the free energies of the two phases,
$$e^{-F/k_BT}=Tre^{-H/k_BT},\eqno(43)$$
for the same sets of parameters.
\par
{}From the above figures, it is possible to understand the competition between
the
LTT and LTO phases as a competition between energy and entropy.  First,
consider
the small overlap case, Fig. 5a.  As $S\rightarrow 0$, the LTO phase becomes
two
{\it independent} LTT-type transitions, in wells 1 and 2, so two branches of
the
energy
curve are lowered by virtually the same amount as the single branch in the
LTT phase.  Thus, {\it for a fixed value of $<cos\phi >$}, the LTO phase has a
greater entropy associated with it.  However, this greater entropy in turn
means that {\it at a fixed temperature}, the self-consistent value of $<cos\phi
>$ will be smaller in the LTO phase than in the LTT phase, Eq. 41.  Hence, for
small $S$ values, the LTT phase is always energetically preferred.
\par
As $S$ increases, however, the factor of level repulsion adds a further
stabilization to the LTO phase (Fig 5b-d).  The splitting of the originally
degenerate pair of LTO levels drives one lower in energy than the corresponding
LTT level.  Also, the mixing of different wells brings two LTT levels fairly
close together, so the entropy differences between the LTO and LTT phases are
reduced.  The result is that both phases are close in free energy over most of
the temperature range, Fig. 8, and that both phases allow non-zero solutions
for
$<cos\phi >$ out to significantly higher temperatures as $S$ increases.
However, the LTO phase wins out at the higher temperatures, and over a wider
temperature range as $S$ gets larger -- the phase diagram is illustrated in
Fig. 9.  Other choices for the parameters $\bar H$ and $H_0^{\prime}$ would
mainly shift the scales of the phase diagram, without altering its fundamental
character.  Note that the LTT phase never increases its transition temperature
significantly above the $S\rightarrow 0$ limit, whereas the LTO phase turns on
at higher T's with increasing $S$.
\par
While the above behavior is qualitatively what might be expected, it should be
cautioned that the above calculations are perturbative in $S$, so the detailed
nature of the large $S$ results should be treated with caution (for instance,
the LTO transition temperature appears to diverge as $S^2\rightarrow 1/2$).
A more detailed calculation would require the inclusion of more distant
overlaps
in Eq. 39.  However, the reduction of the Mathieu problem to a four-state
problem remains valid, as long as $4\alpha$ is large compared to $k_BT$.
Moreover, as Fig. 9
illustrates, for some parameter choices the LTO phase remains stable down to
small $S$-values.
\par
{}From Fig. 7, the transition from the disordered JT phase (HTT) to the LTT
phase is second order, with $<cos\phi >$ acting as an order parameter, but the
LTO$\rightarrow$HTT and LTO$\rightarrow$LTT phase transitions are first order,
with discontinuous jumps in $<cos\phi >$.
Note that in the LTO phase, for $S\ge 0.1$, $<cos\phi >\sim 0.4$ at low
temperatures -- i.e., the macroscopic average tilt is noticeably smaller than
its instantaneous value, $\sqrt{<cos\phi >^2+<sin\phi >^2}\simeq 0.56$.
\par
While the present dynamic JT calculation has been carried out for the
high-T$_c$
cuprates, it is interesting to compare it with calculations of the multiple
phase transitions in the ferroelectric perovskite BaTiO$_3$[38].  There, a
similar competition between energy and entropy was found, with the largest
energy lowering associated with the static (rhombohedral) JT phase, and
increasingly large entropy contributions in the case of two, four, or eight
potential minima being involved in the dynamic JT effect.  However, in the
simplest calculation, all four phases had a transition at the same temperature,
so the static JT phase was stable at all temperatures.  By assuming that the
interaction energy was different in the different phases, it was possible to
reproduce the observed sequence of phase transitions.  In the present
calculation, the overlap parameter naturally provides the stabilization for the
dynamic phases, with only the static phase stable as $S\rightarrow 0$.
\bigskip
\noindent{\bf 8. Discussion}
\smallskip\par
\bigskip
\noindent{\bf 8a. Dynamic JT}
\smallskip\par
Figure 9 is the chief result of this paper.  It confirms the suggestion[10,11]
that the LTO phase can be described as a dynamic JT phase involving a splitting
of the vHs degeneracy.
\par
The present paper has contrasted two possible origins of the LTO phase, as a
static JT phase stabilized by umklapp scattering (Section 5), and as a dynamic
JT phase, Section 7c, which is naturally related to the LTT phase.
By comparing the gap functions of the two models, Figs. 2 vs 7a, it can be seen
that, for small $S$ values, the dynamic JT effect can mimic the effect of a
static transition driven by umklapp scattering, thereby confirming the
speculation made in  Paper VIIIA[10].  For larger $S$ values, the situation is
more complicated, with the LTO transition becoming first order.
\par
Much work remains to be done, particularly in simultaneously accounting for
both the electronic and structural aspects of the transition, and in
understanding how electronic properties are modified within the dynamic JT
phase
(antiferromagnetism, superconductivity, ...).  Before this can be done,
however,
some more fundamental questions must be answered, such as, what does the
Fermi surface (or even the Brillouin zone) {\it mean} in a dynamic JT system,
where the local symmetry is not the same as the global symmetry, and indeed
where the local symmetry can fluctuate in space and time.
\bigskip
\noindent{\bf 8b. SDW-CDW Competition Revisited}
\smallskip\par
This paper has explored possible generalizations of CDW's and SDW's in the
presence of a vHs.  While the simple CDW can describe the LTT phase, it seems
to
be ruled out by a symmetry argument in the LTO phase (Ref. [9] and Appendix
III).  Nevertheless, there are at least two
mechanisms by which vHs-related effects could stabilize an LTO phase.  Which of
these mechanisms is actually operative in LSCO is a question which requires
considerably more research.  Nevertheless, it seems appropriate to point out
that even at the molecular level, there is often a competition between JT and
spin-orbit effects[12,13], and this competition is reminiscent of the usual
CDW-spin-density wave (SDW) competition in lower-dimensional metals in the
presence of a peak in the dos.
\par
Near half filling, correlation effects become important in destabilizing
CDW-like phases.  On the other hand, there has been
clear experimental[39] and theoretical[40] evidence for the importance of
spin-orbit coupling in the undoped materials, and specifically in the
N\'eel antiferromagnetic or spiral magnetic phases.  This can be understood,
in the context of the present paper, as follows.  To split the vHs degeneracy
requires breaking the degeneracy of the two Cu's in the orthorhombic unit cell.
This can be done if they have opposite spins, but this requires spin-orbit
coupling to modify the electronic bands.
\par
I earlier suggested that the phase diagram of LSCO resembled a crossover from
SDW-like behavior near half filling (the antiferromagnetic state) to
`incipient CDW-like' behavior, including Peierls distortion, as hole doping is
increased[28].  The present results suggest the following modification:
the hole-doped LTO phase is stabilized by the dynamic JT effect, whereas
near half filling spin-orbit effects are more important (due to on-site
Coulomb repulsion), and may stabilize a static LTO phase.  This phase may be
related to the proposed `flux phase'[18].  Whereas in principle,
a single spin-orbit coupled phase could persist for all dopings, there is
considerable experimental evidence for a transition between two phases as a
function of doping, with a (nanoscale) two-phase regime between the vHs and
half filling[28].
\par
In Section 2, it was briefly noted that c-axis dispersion acted in the same way
as a CDW to split the vHs degeneracy.  It would be interesting to study in
more detail whether changes in interlayer hopping could drive a phase
transition, particularly in light of Anderson's ideas that interlayer
coupling plays a special role in stabilizing high-T$_c$ superconductivity[41].
\bigskip
\noindent{\bf 9. `Excitons' vs `Cageons'}
\smallskip\par
When I initiated this series of calculations on the vHs[42], a major premise
was that inter-vHs scattering could promote large electron-phonon coupling
via vHs nesting [N.B. not conventional nesting], leading to short-range CDW
order.  Indeed, the term `exciton' was introduced to point out that the
strong scattering is associated with electron-like sections of one vHs
scattering off of hole-like sections of the second vHs.  The CDW order in
this case would be an `excitonic instability' analogous to the spin-density
wave instability of chromium.
\par
The `excitonic' properties of the model were discussed in paper IV[43]; the
CDW in V[27].  The present series of papers, VIIIA-C, are an extension of V:
there is a structural instability, but it is not simply describable as a CDW
(the formalism of V is still relevant for extending the
present model to a quasi-two-dimensional system and incorporating mode-mode
coupling).
\par
Under these circumstances, the term `excitonic' does not seem to be
particularly suitable for describing the system, since the strong
electron-phonon coupling will lead to excitations closer to polarons -- or
even solitons, as discussed below.  Hence, I am introducing the term `cageon',
which is intended to better describe the JT excitations of electrons coupled to
tilts of the octahedral CuO$_6$ cages.
\par
Going beyond the mean field calculation, the corner sharing of the CuO$_6$
octahedra suggests that there should be long chains of LTT phase, and that
defects must be introduced into the chains.  A plausible model would be to
have islands of LTT phase separated by LTO-like domain walls, which would
switch an x-directed domain into a y-directed one.  Such domain walls would
behave as solitons, and the doping dependence of LBCO could be interpreted
in terms of the generation of these solitons.  Thus, at $x=0.12$ ($6\%$ of the
La replaced by Ba), the material is in a pure LTT phase.  As x is reduced,
the material transforms to the LTO phase.  This transformation could be
accomplished via soliton generation -- as x is decreased, the density of LTO
solitons increases.  The octahedral shear $e_{12}$ would simply be proportional
to the soliton density.  This idea will be pursued further in a future
publication.
\par
In particular, it has been found that the degeneracy of the vHs's is split in
the solitonic model.  It is just this splitting which underlies the `valence
bond density wave' calculations of Ref. [10].  Hence, it seems likely that the
results of that paper hold for the dynamic JT model of the LTO phase -- in
particular, the phase diagram of the transitions $HTT\rightarrow LTO\rightarrow
$ superconductor should continue to hold in the dynamic JT model.
\par
I would like to thank J. Zak for explaining the role of the Berry phase in the
JT effect.  Publication 545 from the Barnett Institute.
\bigskip
\centerline{\bf Appendix I: Renormalized Microscopic JT Band Structure}
\bigskip
\par
The present analysis, in terms of $\rho_{\pm}$, is very convenient, but how
can it be related to a microscopic Hamiltonian, such as that developed in
VIIIB,
Appendix I?  In particular, the analysis of Section 5 requires being able to
define $a_1^{\dagger}(\vec l)$, $a_2^{\dagger}(\vec l)$ -- i.e., on each atomic
site.  By contrast, the microscopic Hamiltonian involves interatomic hopping
between Cu and O atoms.  In this Appendix, I show that a site model can be
derived from the microscopic Hamiltonian, if the sites are not individual
atoms, but {\it clusters} of atoms.
\par
The simplest cluster is a single octahedron -- or more simply, a Cu atom with
the four surrounding planar O's, since the present model does not incorporate
either the Cu $d_{z^2}$ nor the apical O p orbitals.  However, one octahedron
is too small.  There is only a single Cu-O antibonding level (twofold
degenerate due to spin) per Cu atom, and hence no JT degeneracy.  Hence, the
appropriate cluster contains a square of {\it four} octahedra.  The antibonding
`band' contains four levels, the middle two of which are degenerate, in the
absence of strain or tilt coupling.  This is a JT degeneracy, since the
antibonding `band' is half filled (the model of VIIIB contains no O-O hopping,
so the vHs falls exactly at half filling).
\par
The dispersion of the four levels can readily be recovered from VIIIB.
$$E={\Delta E\over 2}\pm\sqrt{({\Delta E\over 2})^2+4W},
\eqno(I1a)$$
$$W=t_{CuOx}^2[cos^2\theta_xsin^2({k_xa\over 2})+\beta_{\pi}^2
sin^2\theta_xcos^2({k_xa\over 2})]+
t_{CuOy}^2[cos^2\theta_ysin^2({k_ya\over 2})+\beta_{\pi}^2
sin^2\theta_ycos^2({k_ya\over 2})].\eqno(I1b)$$
Here, $\Delta E$ is the splitting of the Cu and O levels, and $t_{CuOx}$ ($t_{
CuOy}$) and $\theta_x$ ($\theta_y$) are the Cu-O hopping energy and octahedral
tilt along the x (y) axis, respectively.  For a `sample' two cells by two
cells,
$k_xa/2$ and $k_ya/2$ are restricted to the values $0$ and $\pi /2$,
leading to four possible values of $E$, Eq. I1a, with eigenfunctions shown in
Fig. 10 (top).  (Strictly speaking, these
are not small clusters, because periodic boundary conditions are assumed.
Calculations on real clusters would reveal eigenstates of the same overall
symmetry as in Fig. 10, but with modified eigenvalues.)
\par
This same procedure may readily be generalized to larger `samples'.  For a
sample $N\times N$ cells, there are $N^2$ levels in the antibonding band,
of which $N$ states are degenerate (when the strains and tilts are absent) at
the vHs energy level -- those states for which $k_x+k_y=\pi /a$.  For these
states, the energy can be written as Eq. I1, with
$$W=W_a+W_bcos(k_xa),\eqno(I2a)$$
$$W_a=t_{CuO}^2[1-{a\over r^*}e_++{4(\beta_{\pi}^2-1)\over 3ma^2}\bar R^2],
\eqno(I2b)$$
$$W_b=t_{CuO}^2[{a\over r^*}e_-+{4(\beta_{\pi}^2+1)\over 3ma^2}(R_1^2-R_2^2)],
\eqno(I2c)$$
where $r^*$, $\beta_{\pi}$, $R_i^2=3ma^2sin^2\theta_i/8$, and $t_{CuOi}^2=
t_{CuO}^2(1-2ae_{ii}/r^*)$, as discussed in VIIIB (see also the last paragraph
of this appendix).  This equation offers a
convenient {\it microscopic} estimate of the electron-phonon coupling
constants.
\par
In order to find the renormalized bands corresponding to $\rho_{ij}$ of Eq. 4,
it is convenient to look at how larger clusters are built up.
Fig. 10 (bottom) shows the eigenfunctions associated with the $4\times 4$
clusters.  Beneath each figure, the corresponding eigenvalue is indicated in
the form $ij$.  For simplicity, this notation corresponds to the untilted
clusters ($\theta_x=\theta_y=0$), for which $W$ in Eq. I1b is given by $4W=i
t_{CuOx}^2+jt_{CuOy}^2$.  The $4\times 4$ eigenfunctions are built up from the
$2\time 2$ solutions by
repeated tiling of the $4\times 4$ cell by the $2\times 2$ cell or its
negative.
Since there is no intercell mixing, the Cu-O band separates into a
superposition
of four overlapping but noninteracting subbands.  This pattern holds for larger
cells, but with some minor complications: (a) all of the '+' Cu atoms need not
have the same amplitude [example: for the one-dimensional chain 8 cells long,
the eigenfunction ($+\ +\ +\ +\ -\ -\ -\ -$) is really $(a,b,b,a,-a,-b,-b,-a)$,
with $a/b=\sqrt{2}-1$]; and (b) most levels are degenerate in pairs, leading to
eigenvalues of more complicated form.  These complications do not affect the
subband separation.
\par
{}From studying larger cells, a separate energy dispersion can be determined
for
each subband.  These energy dispersions have the form of Eq.I1, with a
restricted range of $k_x$, $k_y$, for each subband.  Thus, for subband I (IV),
$k_x$ and $k_y$ must both be less than (greater than) $\pi /2$, while for
subband II (III), $k_x$ ($k_y$) is greater, while $k_y$ ($k_x$) is less than
$\pi /2$.  This restriction can most elegantly be carried out by introducing
new {\it artificial Brillouin zone boundaries}, as illustrated by the dashed
lines in the inset to
Fig. 11.  For the case of untilted molecules, the subband dispersions are
illustrated in Fig. 11.  It can be seen that subbands II and III are degenerate
(in the untilted case), and overlap the vHs, while band I(IV) lies
below (resp. above) the vHs.  For the purposes of the present calculation,
bands
I and IV may be neglected, while bands II and III correspond to the electronic
bands discussed in the text, e.g., Eq. 4.  In particular, since these bands are
asymmetric in $k_x$ vs $k_y$, their degeneracy will be lifted by a nonvanishing
LTT-type tilt.
\par
It should be noted that Schulz[17] has introduced a similar formalism with
(spin-dependent) creation operators associated with each vHs.
\par
{\bf ADDENDUM to Paper VIIIB[11]}.  In Paper VIIIB, the covalent overlap
enhancement factor $\beta_{\pi}$ was introduced as
$$\beta_{\pi}=1+{2\tau_{\pi}\over\sqrt{3}\tau_{\sigma}}.$$
Here $\tau_{\sigma}$ and $\tau_{\pi}$ are related to the Slater-Koster[44]
parameters of p-d overlap of $\sigma$ or $\pi$ symmetry as follows:
$\tau_{\sigma}=V_{pd\sigma}$, $\tau_{\pi}=-V_{pd\pi}$ (N.B., $V_{pd\pi}$ is
negative).  At the time of writing, I was unable to find theoretical values for
the two overlaps separately.  Now Grant and McMahan[45] have provided {\it ab
initio} calculations for tetragonal La$_2$CuO$_4$ of sufficient detail to allow
an estimate of these parameters.
Their calculations provide two estimates of $\tau_{\sigma}$.  From
McMahan and Grant, Table I: $t(d_{x^2-y^2},p_{\sigma})=\sqrt{3}\tau_{\sigma}/2=
1.47eV$; $t(d_{3z^2-r^2},p_{\sigma})=\tau_{\sigma}=0.50eV$.  These estimates
are
not exactly equal, since the wave functions are Wannier functions which are not
pure $p$ and $d$ states, due to overlap with higher orbitals.  From Grant,
Table
2.5 $t(d_{xy},p_{\pi})=\tau_{\pi}=0.72 eV$.  Averaging the $\tau_{\sigma}$
estimates, this leads to a theoretical value for $\beta_{\pi}\simeq 1.6$,
somewhat smaller than the value used in Ref. [11].  This value is in good
agreement with recent cluster calculations[46]: $V_{pd\sigma}=1.5eV$, $V_{pd
\pi}=-0.7eV$, yielding $\beta_{\pi}=1.5$.  I would like to thank A.
McMahan for providing me with copies of Ref. [45].
\bigskip
\centerline{\bf Appendix II: Phonon-mediated Electron-electron Interaction}
\bigskip
\par
The strain terms can be decoupled from the Hamiltonian, Eq. 1, by defining
$$\tilde e_{\pm}(q)=e_{\pm}(q)+{G_{\pm}\rho_{\pm}(q)+F_{\pm}{\cal
R}_{\pm}^{(3)}
({q\over 2})\over 2C_{\pm}},\eqno(II1a)$$
$$\tilde e_{12}(q)=e_{12}(q)+{2G_{66}\rho_{12}(q)+
F_{66}{\cal R}_0^{(3)}({q\over 2})\over 4C_{66}}.\eqno(II1b)$$
The reduced Hamiltonian becomes $H=H_{str}^{\prime}+H_+^{\prime}+H_-^{\prime}+
H_0^{\prime}+H_{ph}+H_e$, with
$$H_{str}^{\prime}=\sum_q\Bigl(C_+\tilde e_+(-q)\tilde e_+(q)+
C_-\tilde e_-(-q)\tilde e_-(q)+2C_{66}\tilde
e_{12}(-q)\tilde e_{12}(q)\Bigr),\eqno(II2a)$$
$$H_{\pm}^{\prime}=\sum_q\Bigl({\cal R}_{\pm}^{(3)}({q\over 2})
\bigr(\Gamma_{0\pm}^{\prime}{\cal R}_{\pm}^{(3)}({-q\over 2})
+\tilde\alpha_{\pm}^{e\prime}\rho_{\pm}(-q)\bigr)\Bigr)
-\sum_q{J_{\pm}^{\prime}\over 2}\rho_{\pm}(-q)\rho_{\pm}(q),\eqno(II2b)$$
$$H_0^{\prime}=\sum_q\Bigl({\cal R}_0^{(3)}({q\over 2})
\bigr({\Gamma_2^{\prime}\over 2}{\cal R}_0^{(3)}({-q\over 2})
+\tilde\gamma^{e\prime}\rho_{12}(-q)\bigr)\Bigr)
-\sum_q{J_0^{\prime}\over 2}\rho_{12}(-q)\rho_{12}(q),\eqno(II2c)$$
where $\Gamma_{0\pm}^{\prime}=-F_{\pm}^2/2C_{\pm}$, $\Gamma_2^{\prime}=
-F_{66}^2/16C_{66}$, $\tilde\alpha_{\pm}^{e\prime}=\tilde\alpha_{\pm}^e-G_{\pm}
F_{\pm}/2C_{\pm}$,
$\tilde\gamma^{e\prime}=\tilde\gamma^e-G_{66}F_{66}/4C_{66}$,
$J_{\pm}^{\prime}=G_{\pm}^2/2C_{\pm}$, and $J_0^{\prime}=G_{66}^2/2C_{66}$.
The term $H_{str}^{\prime}$ is decoupled from the remaining terms, and since it
is purely quadratic, does not contain any interesting dynamics, hence can be
neglected.
\par
The form of $H_{ph}$, Eq. 5d, does not allow a similar decoupling of the tilt
terms.  Such a separation can be obtained for a related `toy' Hamiltonian,
which
consists of Eq. 1, with $H_{ph}$ replaced by
$$H_{ph}^T={\omega^{2\prime}\over 2}{\cal R}_+^{(3)}(0)+{1\over 2}
\sum_q\Bigl(\tilde\Gamma_0[{\cal
R}_+^{(3)}({q\over 2}){\cal R}_+^{(3)}({-q\over 2})+
{\cal R}_-^{(3)}({q\over 2}){\cal R}_-^{(3)}({-q\over 2})]
+{\tilde\Gamma_2\over 2}{\cal R}_0^{(3)}({q\over 2})
{\cal R}_0^{(3)}({-q\over 2})\Bigr).\eqno(II3a)$$
If $\tilde\Gamma_0=(\Gamma_0+\Gamma_0^a)/16$, $\tilde\Gamma_2=\Gamma_2/16$, and
$\omega^{2\prime}=\tilde\omega^2(q_R)$, then
Eq. II3 will be identical to Eq. 5d for the soft mode, $q=0$, $\vec q^{\prime}=
\vec q_R\equiv (\pi /a,0)$ or $(0,\pi /a)$.
Equation II3 can now be incorporated into Eqs.
II2b,c by substituting $\Gamma_{0\pm}^{\prime}\rightarrow\Gamma_{0\pm}^{\prime
\prime}\equiv\tilde\Gamma_0+\Gamma_{0\pm}^{\prime}$ and $\Gamma_2^{\prime}
\rightarrow\Gamma_2^{\prime\prime}\equiv\tilde\Gamma_2+\Gamma_2^{\prime}$ into
the latter equations, and replacing $H_{ph}^T\rightarrow H_{ph}^{\prime}$, with
$$H_{ph}^{\prime}=\omega^{2\prime}{\cal R}_+^{(3)}(0).\eqno(II3b)$$
At this point, the tilt couplings can be formally
decoupled from the toy Hamiltonian by a procedure similar to the strain
decoupling, by defining
$$\tilde{\cal R}_{\pm}^{(3)}({q\over 2})={\cal R}_{\pm}^{(3)}({q\over 2})+
{\tilde\alpha_{\pm}^{e\prime}\rho_{\pm}(q)\over 2\Gamma_{0\pm}^{\prime\prime}},
\eqno(II4a)$$
$$\tilde{\cal R}_0^{(3)}({q\over 2})={\cal R}_0^{(3)}({q\over
2})+{2\tilde\gamma
^{e\prime}\rho_{12}(q)\over \Gamma_2^{\prime}}.\eqno(II4b)$$
In this case, $H=H_{str}^{\prime}+H_R+H_{\rho}$, with
$$H_R=\omega^{2\prime}\tilde{\cal R}_+^{(3)}(0)+\sum_q\Bigl(\sum_{i=\pm}[
\Gamma_{0i}^{\prime\prime}\tilde {\cal R}_i^2({-q\over 2})\tilde {\cal R}_i^2
({q\over 2})]+{\Gamma_2^{\prime\prime}\over 2}\tilde {\cal R}_0^2({-q\over 2})
\tilde {\cal R}_0^2({q\over 2})\Bigr),\eqno(II5a)$$
$$H_{\rho}=-\sum_q\Bigl({J_+\over 2}\rho_+(-q)\rho_+(q)+
{J_-\over 2}\rho_-(-q)\rho_-(q)
+{J_0\over 2}\rho_{12}(-q)\rho_{12}(q)\Bigr),\eqno(II5b)$$
with
$$J_{\pm}={G_{\pm}^2\over 2C_{\pm}}+{\tilde\alpha_{\pm}^{e\prime 2}\over 2
\Gamma_{0\pm}^{\prime\prime}}\eqno(II6a)$$
and
$$J_0={G_{66}^2\over 2C_{66}}+{4\tilde\gamma^{e\prime 2}\over\Gamma_2^{\prime
\prime}}.\eqno(II6b)$$
Even for the toy model, this separation must be treated with caution, because
of
the unusual form of the substitution, Eq. II4.  In the strain case, the
separation can be carried out by a canonical transformation, whereas in the
present problem, this does not seem to be the case, due to the quadratic term
in
$R$ in Eq. II4.
\bigskip
\centerline{\bf Appendix III: Structural Instabilities in the Uniform LTO
Phase}
\bigskip
\par
This Appendix analyzes the question of whether a uniform LTO phase could be
brought about by electron-phonon interaction.  That is, can the large dos
associated with the vHs's be driven below the Fermi level by a lattice
distortion.  It has been shown that the LTO transition does not split the
degeneracy of the vHs's[9], but in itself, this is not sufficient.  There are
two alternative means by which the dos could be shifted below the Fermi level,
without splitting the vHs degeneracy.  Here it is shown that neither of these
effects arises in LSCO, in the absence of spin-orbit interaction.
\par
These effects are: (1) there is a single peak in the dos, due to both
(degenerate) vHs's, but the structural transition shifts this peak below the
Fermi level; or (2) umklapp scattering couples two vHs's opening a gap in the
dos, as discussed in Section 5.  Case (1), which will be discussed in Section
IIIa, would arise if, for example, due to the orthorhombic distortion, the
vHs's
no longer occured at the corners of the orthorhombic Brillouin zone, or if the
transition reduced the magnitude of the average O-O hopping matrix element.
\bigskip
\centerline{\bf IIIa: Orthorhombic Distortion}
\bigskip
The Brillouin zone of Fig. 1b for the LTO phase
has been oversimplified by the neglect of the orthorhombic splitting.  The LTO
phase modifies the Brillouin zone of the HTT tetragonal phase in two ways.
First, the principal axes $a^*$ and $b^*$ are rotated by $45^o$ with respect to
the tetragonal axes, $a$ and $b$, Fig. 1b.  (The real
space cell is doubled in area, so the Brillouin zone is halved.)  Secondly,
there is a small orthorhombic distortion, $a^*\ne b^*$, which was neglected in
Fig. 1b.
\par
This distortion can readily be incorporated into the tight-binding calculations
of Appendix I.  When these calculations are repeated for the larger LTO unit
cell, two changes arise.  First, each band becomes two-fold degenerate (due to
the two Cu's per unit cell).  Secondly, in all dispersion relations, such as
Eqs. I1, I2, the following substitutions must be made:
$$k_xa\rightarrow{\delta_++\delta_-\over 2},\eqno(III1a)$$
$$k_ya\rightarrow{\delta_+-\delta_-\over 2},\eqno(III1b)$$
with $\delta_+=k_+a^*$, $\delta_-=k_-b^*$, and $k_{\pm}$ the wave vectors along
the new principal axes.  As a result of this transformation, the topology of
the
Fermi surfaces is maintained as the Brillouin zone is stretched into a
rectangle.  In particular, the vHs's continue to intersect the Brillouin zone
boundaries in the corners of the zone, as in Fig. 1b.  Thus, the orthorhombic
distortion does not alter the ratio of the area of a given Fermi surface to the
total Brillouin zone area -- and hence does not change the doping at which
the vHs's coincide with the Fermi level.
\par
This {\it still} does not rule out the possibility of
a shift of the vHs away from the Fermi level.  Thus, the doping $x=x_c$ at
which the vHs coincides with the Fermi surface is controlled by the curvature
of
the Fermi surface at the vHs.  The ratio of the area of the hole Fermi surface
to the total Brillouin zone area is $(1+x_c)/2$.  But the curvature of the
Fermi
surface is proportional to the O-O hopping parameter, $t_{OO}$: when
$t_{OO}=0$,
the Fermi surface is square.  Since the orthorhombic distortion changes all
O-O distances (and produces two inequivalent distances), it could lead to
a change in the average value of $t_{OO}$, and hence a shift of $x_c$.  Such an
effect is likely to be small, since one O-O separation increases while the
other
decreases, so the corresponding changes in $t_{OO}$ tend to cancel.
\bigskip
\centerline{\bf IIIb: Umklapp Scattering and Group Theory}
\bigskip\par
The role of umklapp scattering can best be appreciated by analyzing a Fermi
surface away from the vHs, on the side of overdoping, Fig. 12a.  Simple zone
folding from the tetragonal to the orthorhombic Brillouin zone produces
the lens-shaped orbits shown in Fig. 12a.  However, these orbits are anomalous,
having discontinuous slopes at the Brillouin zone boundary.  In most band
structures, umklapp scattering causes an interaction between carriers from
opposite sides of the Brillouin zone, opening a gap between successive
bands.  This causes the Fermi surfaces of each band to intersect the Brillouin
zone boundaries orthogonally, so that in crossing the zone boundary (in an
extended-zone scheme) the carrier stays within the same band, and there is no
slope discontinuity.
\par
Such an interaction will arise in a tight-binding model in the following
fashion.  The orthorhombic unit cell contains twice as many atoms as the
original tetragonal cell.  This will double the size of the energy eigenvalue
matrix, and the number of bands, leaving the bands degenerate in pairs.  To
split the degeneracy of the
bands requires that the matrix elements of formerly equivalent atoms (e.g., the
two Cu atoms in the cell) be different.  However, there is a (glide) symmetry
operation
which can translate one Cu into the other, so the matrix elements can only
differ in overall sign -- e.g., terms proportional to $\pm sin\theta$, in
Appendix I.  Since these terms enter the eigenvalue equation only in the
square, they cannot produce an energy splitting.  Thus, umklapp coupling
is absent -- $\rho_{12}$=0.
\par
But how can the slope discontinuities of the lens orbits be accounted for?
This can best be understood by a general symmetry argument, based on the $Bmab$
space group of the LTO phase.  There is a glide plane perpendicular
to the $b^*$ direction, which interchages the two Cu atoms in a cell.  Because
of this, {\it all} eigenfunctions on the $a^*$-face of the Brillouin zone must
be {\it doubly degenerate}[47].  In the presence of such degeneracy, the slope
of the $E(k)$ curves can be discontinuous.  Such a situation always arises in
the presence of glide planes, and is perhaps best known for the hexagonal close
packed space group[48].  In this case, one zone boundary does not introduce a
gap, and the $E(k)$ curve from the first band merges continuously into the
second band on crossing the zone boundary.  It is often convenient to ignore
the
zone boundary, and work with a larger k-space zone[49].  Such a double zone
for the LTO phase is illustrated in Fig. 12b: the zone is doubled in the $X^*$
direction (along $a^*$), while a gap is allowed along the $Y^*$ direction.
Thus, lens orbits appear only along the zone face at $Y^*$, with open orbits
along $X^*$[50].
\bigskip
\centerline{\bf IIIc: Spin-Orbit Interaction}
\bigskip
\par
Spin-orbit coupling lifts most of these degeneracies, so that the enlarged
k-space zone can no longer be used[51].  The $Bmab$ space group is equivalent
to the $Cmca$ space group, which includes the structure of the elements Br, I,
and Ga.  The group theoretical analysis for Ga has been carried out by
Koster[52], who finds that spin-orbit coupling eliminates most of the
degeneracy
on the faces of the Brillouin zone.  This is illustrated in Fig. 13.  A
complication arises in that the orthorhombic unit cell, with axes $a^*$, $b^*$,
and $c$ is not primitive.  Figure 13a shows this cell, with the primitive cell
(containing only half as many atoms) inscribed in it.  The Figure also shows
the Brillouin zones corresponding to the primitive cell (Fig. 13b) and the
orthorhombic cell (Fig. 13c) -- for convenience, the latter zone will be called
the pseudozone.  Its importance arises because it is the natural zone for
tight-binding model calculations, particularly in the two-dimensional limit
when
the energy bands are assumed independent of $k_z$.  Figure 13 illustrates how
the true zone may be folded into the pseudozone.  Figure 13b shows the first
pseudozone inscribed in the full zone, while Fig. 13d shows how the second
pseudozone is reassembled from the leftover parts of the full zone.  This
folding produces two bands, which are degenerate within a tightbinding
calculation such as that of Appendix I.  Note the relative orientation of the
two pseudozones in the full zone: the second zone is predominantly displaced
from the first by the $Q$-vector ${\vec Z^*}={\vec Z}/2$.  This is because the
unit cell (Fig. 13a) includes contributions from two CuO$_2$ planes displaced
along the c-axis, which are equivalent in the tight-binding scheme.
\par
By including additional terms in the tight-binding calculation, it is in
principle possible to couple the two bands and remove the twofold
degeneracy.  However, a group theoretical analysis shows that not all of the
degeneracy can be lifted.  In Figs. 13b,c, the
hatched areas show the regions in the Brillouin zone (all confined to the
surface of the zone) in which the wave functions are two-fold degenerate
(neglecting spin degeneracy) in the absence of spin-orbit coupling: this
happens throughout the $L-X-N$ plane and along the line $M-N$.  When
spin-orbit coupling is included, the degeneracy is lifted except on the
regions which are heavily shaded: along the $X-L$ line and at the $M$ point.
\par
\par
{}From the pseudozone of Fig. 13c, the
approximate two-dimensional Brillouin zone (Figs. 1b, 12a) is found by
neglecting the band dispersion along the c-axis, which should be a good
approximation for the cuprates.  From Figure 12, it can be seen that spin-orbit
coupling lifts the degeneracy over must of the Brillouin zone boundary.  In
particular, the lens orbits will be continuous at the zone boundary, due to the
opening of a gap, and spin-orbit coupling can lead to an umklapp gap
even at the vHs, thereby stabilizing a uniform LTO phase, as discussed in
Section 5.
\bigskip
\centerline{\bf Appendix IV. Pseudospin Formalism}
\smallskip\par
\par
The calculations of Sections 6b,c are similar to previous studies of
structural phase transitions[12-14,29], which have introduced pseudospin
formalisms, to describe either the electronic or the phonon modes.  These
formalisms are often based on an imperfect analogy, but are valuable because
spin systems are understood so much better than nonlinear phonon systems.
To clarify comparison with these works, the formalism is discussed
in this Appendix.  The pseudospin formalism applies to the order-disorder limit
of structural phase transitions.  That is, it is
assumed that at each lattice site the system is disordered into one of two
(or more) possible configurations.  In the present problem, an electronic
pseudospin can be introduced, with the `up' spin associated with occupancy
of the $\rho_{11}$ vHs, and the `down' spin with $\rho_{22}$ electrons.  Then
$$\sigma_{zi}=a^{\dagger}_{1i}a_{1i}-a^{\dagger}_{2i}a_{2i},\eqno(IV1a)$$
$$\sigma_{xi}=a^{\dagger}_{1i}a_{2i}+a^{\dagger}_{2i}a_{1i},\eqno(IV1b)$$
$$\sigma_{yi}=i(a^{\dagger}_{2i}a_{1i}-a^{\dagger}_{1i}a_{2i}),\eqno(IV1c)$$
where the subscript $i$ labels the atomic site.
The spin operators satisfy the commutation relations
$$[\sigma_{il},\sigma_{jm}]=2i\epsilon_{ijk}\delta_{lm}\sigma_{kl}.\eqno(IV2)$$
\par
The $\rho$ operators can then be rewritten in terms of pseudospins.  In
particular,
$$\rho_-(0)=\sum_k\bigl(a_{1k}^{\dagger}a_{1k}-
a_{2k}^{\dagger}a_{2k}\bigr)=\sum_i\sigma_{zi}\equiv S_z,\eqno(IV3a)$$
$$\rho_-(q)=\sum_i\sigma_{zi}e^{i\vec q\cdot\vec r_i}\equiv S_z(q).
\eqno(IV3b)$$
Thus, condensation into the LTT phase corresponds to a pseudospin {\it
ferromagnetic} transition within the plane, $<S_z>\ne 0$; the overall
tetragonal symmetry arises because the interlayer coupling is
antiferromagnetic.
\par
In the present model, the pseudospins are again only of approximate validity.
The problem lies in the term $\sum_{ik}E_ka^{\dagger}_
{ik}a_{ik}$.  In an order-disorder model, each electronic state would have to
be either in state 1, associated with the X-point vHs, or state 2, associated
with the Y-point vHs, in which case $a^{\dagger}_{1k}a_{1k}+a^{\dagger}_{2k}
a_{2k}=1$ for each $k$-value.  In fact, while the dos peaks are split, there
remains a significant overlap of the two bands, and for these states, the sum
can reach a value 2.  Hence, the $E_k$-term must be retained in Eq. 19a,
leading
to a very different form of gap equation from a pseudospin calculation.
\bigskip
\centerline{\bf Appendix V: Solutions of Mathieu's Equation (Eq. 32a)}
\smallskip\par
The solutions of Eq. 32a can be written in the form of a Fourier series[31]
$$\Theta =\sum_{m=-\infty}^{\infty}a_me^{im\theta},\eqno(V1)$$
where the boundary conditions require that $m+1/2$ is an integer.  Substituting
V1 into Eq. 32a yields the recursion relation
$$a_m({\cal E}_m-\alpha m^2)+{1\over 2}\beta (a_{m-4}+a_{m+4})=0,\eqno(V2)$$
where ${\cal E}_m$ is the eigenvalue and $\alpha =\hbar^2/2\bar R^2$.  Eq. V2
only mixes $m$-values separated by $\pm 4$, so there are four families of
solution,
depending on whether the series V1 contains terms with $m$ equal to 3/2, 1/2,
-1/2, or -3/2.  Since the recursion relation contains only $m^2$, these terms
are degenerate in pairs for any value of $\beta$.  When $\beta =0$, each term
in V1 is an exact solution, with eigenvalue ${\cal E}_m=\alpha m^2$.  Nonzero
$\beta$ leads to mixing of these states, and for large $\beta$, each level
is fourfold degenerate (corresponding to independent oscillations about one of
the four potential minima).
\par
In finding the eigenvalues of Eq. 32a, it is convenient to generalize the
equation to:
$${d^2y\over dz^{\prime 2}}+(a^{\prime}-2q^{\prime}cos(lz^{\prime}))y=0,\eqno
(V3)$$
where the number of minima, $l$, is arbitrary, and the boundary condition is
that $y(z^{\prime}+l\pi)=-y(z^{\prime})$.  This agrees with Eq. 32a when
$a^{\prime}=E_n/\alpha$, $q^{\prime}=\beta /2$, $z^{\prime}=\phi$, and $l=4$.
In turn, Eq. V3 can be reduced to the canonical
form of a Mathieu function of fractional order[33] by the substitution
$z=lz^{\prime}/2$, $a=4a^{\prime}/l^2$, $q=4q^{\prime}/l^2$, so
$${d^2y\over dz^2}+(a-2qcos(2z))y=0.\eqno(V4)$$
{}From the boundary condition, the solution can be written in the form of
series
V1, with $m=n+p/l$, where $n$ is an arbitrary integer and $p$ is an odd
integer
$\le l$.  These solutions are the fractional Mathieu functions of order $p/l$,
and the eigenvalues can be read off of Fig. 11 of McLachlan[33] (p. 98).
For fixed $n$, there are $l$ solutions lying between the solutions of the
integral Mathieu functions, of orders $n$ and $n+1$ (actually, in the `stable'
zone between the solutions $a_n$ and $b_{n+1}$) -- see Fig. 14 for $l=3,4,5$.
For even $l$, the solutions are all doubly degenerate, while for odd
$l$, there is an additional singly degenerate level whenever $p=l$.  As $l$
increases, the allowed states fill the `stable' intervals of Mathieu's
equation,
with gaps between successive $n$ values, much as the band structure of a
one-dimensional metal fills in as more and more atoms are added to the chain.
\par
Specializing now to the case of interest, $l=4$, the eigenvalues are shown
in Fig. 14, constructed by interpolation from McLachlan's Fig. 11.  The nature
of the eigenstates can be appreciated by going to the $q=0$ limit.  In this
case, the eigenstates are $y_{pc}=cos(p\theta /2)$ or $y_{ps}=sin(p\theta /2)$,
with $p=1,3$.  Thus, $y_{pc}^2=$ 1, 0.5, 0, or 0.5 in the wells 1, 2, 3, and 4,
for either value of $p$.  The difference is that $y_{3c}$ has additional
maxima outside of the potential minima.  The functions $y_{ps}$ are similar,
with wells 1 and 3 interchanged.  Since $y_{1c}$ and $y_{1s}$ are degenerate,
a number of alternative solutions can be constructed, including ones centered
on
wells 2 or 4.  As $q$ increases, the interwell tunneling probability decreases,
and for sufficiently large barriers, the wave functions should localize within
a
single well.  It might be anticipated that the individual wave functions would
narrow, causing the overlap of wave functions between wells to decrease.  A
plausible measure of this decrease would be $\eta_p\equiv y_{pc}^2(2)/y_{pc}^2
(1)$ -- i.e., the overlap probability of the wave function being found at the
center of well 2 given that its peak value is centered in a neighboring well,
1.
Surprisingly, $\eta_p=0.5$, independent of $q$.  Localization arises from
the mixing of the $p=3$ states into the $p=1$ states, so that the gap $\Delta
E$
between these two levels can be taken as a delocalization energy.
\par
The eigenvalues may be found more precisely (Fig. 3) by deriving a
continued-fraction eigenvalue equation.  Letting $v_{np}=a_m/a_{m-4}$, with $m=
(p+8n)/2$, Eq. V2 can be rewritten
$$v_{np}={q\over a-(p+8n)^2-qv_{n+1,p}}.\eqno(V5)$$
Here $p$ is a positive integer which labels the four series, $p=1,3,5,7$.
Equation V5 is readily solved as a continued fraction
$$v_{np}={q\over\displaystyle a-[p+8n]^2-
            {\strut q^2\over\displaystyle a-[p+8(n+1)]^2-
            {\strut q^2\over\displaystyle a-[p+8(n+2)]^2-...}}}.\eqno(V6)$$
Equations V2 and consequently V6 hold for $n\ge 1$, but the equation for $n=0$
is more complicated, since it mixes the series for $p$ and $\tilde p\equiv
8-p$.
This equation can be written in the form
$$v_{1p}={a-q\gamma-p^2\over q},\eqno(V7a)$$
$$v_{1\tilde p}={a-q/\gamma-\tilde p^2\over q},\eqno(V7b)$$
with $\gamma=a_{\tilde p/2}/a_{p/2}$.  Equating the right-hand side of Eq. 6,
for $n=1$, to the right-hand side of Eq. 7 gives the eigenvalue equation for
the Mathieu function -- actually a pair of equations for $v_{1p}$ and
$v_{1\tilde p}$.  The quantity $\gamma$ can be found by subtracting the two
equations.  Letting
$$y={p^2-\tilde p^2+q(v_{1p}-v_{1\tilde p})\over 2q},$$
then
$$\gamma_{\pm}=-y\pm\sqrt{1+y^2}.$$
The two possible $\gamma$ values give the two degenerate eigenstates for each
energy.  Substituting either into Eq. V7a yields a single eigenvalue equation
for $a(q)$, which is solved numerically to generate Fig. 3.
\bigskip
\centerline{\bf References}
\bigskip
\smallskip
\item{[1]}J.D. Axe, A.H. Moudden, D. Hohlwein, D.E. Cox, K.M. Mohanty, A.R.
Moodenbaugh, and Y. Xu, Phys. Rev. Lett. {\bf 62}, 2751 (1989); T. Suzuki
and T. Fujita, J. Phys. Soc. Japan {\bf 58}, 1883 (1990).
\smallskip
\item{[2]}M. Sera, Y. Ando, S. Kondoh, K. Fukuda, M. Sato, I. Watanabe, S.
Nakashima, and K. Kumagai, Sol. St. Commun. {\bf 69}, 851 (1989).
\smallskip
\item{[3]}Y. Koike, T. Kawaguchi, N. Watanabe, T. Noji, and Y. Saito, Sol. St.
Commun. {\bf 79}, 155 (1991);
Y. Maeno, N. Kakehi, M. Kato, and T. Fujita, to be published, Phys.
Rev. B., and Physica C{\bf 185-9}, 909 (1991); M. Sato, N. Sera, S. Shamoto,
and S. Yamagata, Physica C{\bf 185-9}, 905 (1991).
\smallskip
\item{[4]}R.S. Markiewicz, J. Phys. Condens. Matt. {\bf 2}, 6223 (1990);
S. Barisic and J. Zelenko, Sol. St. Commun. {\bf 74}, 367 (1990);
W.E. Pickett, R.E. Cohen, and H. Krakauer, Phys. Rev. Lett. {\bf 67},
228 (1991).
\smallskip
\item{[5]}J. Friedel, J. Phys. Cond. Matt. {\bf 1}, 7757 (1989).
\smallskip
\item{[6]}R.S. Markiewicz, Int. J. Mod. Phys. B{\bf 5}, 2037 (1991).
\smallskip
\item{[7]}D.M. Newns, C.C. Tsuei, P.C. Pattnaik, and C.L. Kane, Comments in
Condensed Matter Physics {\bf 15}, 273 (1992).
\smallskip
\item{[8]}H. Tagaki, R.J. Cava, M. Marezio, B. Batlogg, J.J. Krajewski, W.F.
Peck, Jr., P. Bordet, and D.E. Cox, Phys. Rev. Lett. {\bf 68}, 3777 (1992).
\smallskip
\item{[9]}J.P. Pouget, C. Noguera, and R. Moret, J. Phys. (France) {\bf 49},
375 (1988).
\smallskip
\item{10]}R.S. Markiewicz, Physica C{\bf 193}, 323 (1992).
\smallskip
\item{[11]}R.S. Markiewicz, Physica C{\bf 200}, 65 (1992).
\smallskip
\item{[12]}R. Englman, ``The Jahn-Teller Effect in Molecules and Crystals"
(London, Wiley, 1972); F.S. Ham, in ``Electron Paramagnetic Resonance", ed. by
S. Geschwind (N.Y., Plenum, 1972), p. 1.
\smallskip
\item{[13]}I. Bersuker and V.Z. Polinger, ``Vibronic Interactions in Molecules
and Crystals" (Berlin, Springer, 1989).
\smallskip
\item{[14]}G.A. Gehring and K.A. Gehring, Rep. Prog. Phys. {\bf 38}, 1 (1975).
\smallskip
\item{[15]}J. Labb\'e and J. Friedel, J. Phys. (Paris) {\bf 27}, 153, 303,
708 (1966).
\smallskip
\item{[16]}E. Pytte, Phys. Rev. Lett. {\bf 25}, 1176 (1970); Phys. Rev B{\bf
4}, 1094 (1971).
\smallskip
\item{[17]}H.J. Schulz, Phys. Rev. B{\bf 39}, 2940 (1989).
\smallskip
\item{[18]}G. Kotliar, Phys. Rev. B{\bf 37}, 3664 (1988); I. Affleck and J.B.
Marston, Phys. Rev. B{\bf 37}, 3774 (1988); and T.C. Hsu, J.B. Marston, and I.
Affleck, Phys. Rev. B{\bf 43}, 2866 (1991); P.A. Lee and N. Nagaosa, Phys. Rev.
B{\bf 46}, 5621 (1992).
\smallskip
\item{[19]}E.L. Nagaev, ``Physics of Magnetic Semiconductors" (Mir, Moscow,
1983); V.J. Emery, S.A. Kivelson, and H.Q. Lin, Phys. Rev. Lett. {\bf 64}, 475
(1990).
\smallskip
\item{[20]}K.A. M\"uller, in ``Nonlinearity in Condensed Matter", ed. by A.R.
Bishop, D.K. Campbell, P. Kumar, and S.E. Trullinger (Berlin, Springer, 1987),
p. 234.
\smallskip
\item{[21]}R.E. Cohen, Nature {\bf 358}, 136 (1992).
\smallskip
\item{[22]}E. Pytte and J. Feder, Phys. Rev. {\bf 187}, 1077 (1969); J. Feder
and E. Pytte, Phys. Rev. B{\bf 1}, 4803 (1970).
\smallskip
\item{[23]}A. Bussman-Holder, A. Migliori, Z. Fisk, J.L. Sarrao, R.G. Leisure,
and S.-W. Cheong, Phys. Rev. Lett. {\bf 67}, 512 (1991).
\smallskip
\item{[24]}S. Barisic and I. Batistic, J. Phys. (France) {\bf 49}, 153 (1988).
\smallskip
\item{[25]}C.A. Balseiro and L.M. Falicov, Phys. Rev. B{\bf 20}, 4457 (1979).
\smallskip
\item{[26]}C. Kittel, ``Introduction to Solid State Physics", 5th Ed. (NY,
Wiley, 1976), p. 199.
\smallskip
\item{[27]}R.S. Markiewicz, Physica C{\bf 169}, 63 (1990).
\smallskip
\item{[28]}R.S. Markiewicz, J. Phys. Condens. Matt. {\bf 2}, 665 (1990);
Physica
C{\bf 170}, 29 (1990); and in ``High-Temperature Superconductivity", ed. by
J. Ashkenazi, S.E. Barnes, F. Zuo, G.C. Vezzoli, and B.M. Klein (Plenum, NY,
1991), p. 555.
\smallskip
\item{[29]}R.J. Elliott and A.P. Young, Ferroelectrics {\bf 7}, 23 (1974); L.
Novakovic, ``The Pseudospin Method in Magnetism and Ferroelectricity" (Oxford,
Pergamon, 1975).
\smallskip
\item{[30]}M.D. Kaplan, Physica C{\bf 180}, 351 (1991).
\smallskip
\item{[31]}M.C.M. O'Brien, Proc. Roy. Soc. (London) A{\bf 281}, 323 (1964).
\smallskip
\item{[32]}M. Abramowitz and I.A. Stegun, Eds. ``Handbook of Mathematical
Functions" (Dover, Mineola, NY, 1964).
\smallskip
\item{[33]}N.W. McLachlan, ``Theory and Applications of Mathieu Functions",
(N.Y., Dover, 1964).
\smallskip
\item{[34]}K.-H. H\"ock, G. Schr\"oder, and H. Thomas, Z. Phys. B{\bf 30}, 403
(1978).
\smallskip
\item{[35]}R.B. Potts, Proc. Camb. Phil. Soc. {\bf 48}, 106 (1952); F.Y. Wu,
Rev. Mod. Phys. {\bf 54}, 235 (1982).
\smallskip
\item{[36]}R. Englman and B. Halperin, Phys. Rev. B{\bf 2}, 75 (1970); B.
Halperin and R. Englman, Phys. Rev. B{\bf 3}, 1698 (1971).
\smallskip
\item{[37]}M.V. Berry, Proc. Roy. Soc., London {\bf A 392}, 45 (1984); G.
Delacr\'etaz, E.R. Grant, R.L. Whetten, L. W\"oste, and J.W. Zwanziger, Phys.
Rev. Lett. {\bf 56}, 2598 (1986).
\smallskip
\item{[38]}A.S. Chaves, F.C.S. Barreto, R.A. Nogueira, and B. Z\~eks, Phys.
Rev.
B{\bf 13}, 207 (1976).
\smallskip
\item{[39]}T. Thio, T.R. Thurston, N.W. Preyer, P.J. Picone, M.A. Kastner, H.P.
Jenssen, D.R. Gabbe, C.Y. Chen, R.J. Birgeneau, and A. Aharony, Phys. Rev.
B{\bf
38}, 905 (1988).
\smallskip
\item{[40]}D. Coffey, T.M. Rice, and F.C. Zhang, Phys. Rev. B{\bf 44},
10112 (1991); N.E. Bonesteel, T.M. Rice, and F.C. Zhang, Phys. Rev. Lett. {\bf
68}, 2684 (1992); and  L. Shekhtman, O. Entin-Wohlman, and A. Aharony, Phys.
Rev. Lett. {\bf 69}, 836 (1992).
\smallskip
\item{[41]}J. Wheatley, T.C. Hsu, and P.W. Anderson, Phys. Rev. B{\bf 37}, 5897
(1988).
\smallskip
\item{[42]}R.S. Markiewicz, J. Phys. Condens. Matt. {\bf 1}, 8911 (1989).
\smallskip
\item{[43]}R.S. Markiewicz, Physica C{\bf 168}, 195 (1990).
\smallskip
\item{[44]}J.C. Slater and G.F. Koster, Phys. Rev. {\bf 94}, 1498 (1954); Table
20-1 in W.A. Harrison, ``Electronic Structure and the Properties of Solids"
(Freeman, San Francisco, 1980), p. 481.
\smallskip
\item{[45]}J.B. Grant, Ph.D. Thesis, Lawrence Livermore Nat. Lab. (1991);
J.B. Grant and A.K. McMahan, Phys. Rev. B{\bf 46}, to be published.
\smallskip
\item{[46]}H. Eskes and G.A. Sawatzky, Phys. Rev. B{\bf 43}, 119 (1991).
\smallskip
\item{[47]}R.V. Kasowski, W.Y. Hsu, and F. Herman, Sol. St. Commun. {\bf 63},
1077 (1987).
\smallskip
\item{[47]}H. Jones, ``The Theory of Brillouin Zones and Electronic States in
Crystals", 2d Ed. (North Holland, Amsterdam, 1975), pp. 132-155.
\smallskip
\item{[49]}A.P. Cracknell and K.C. Wong, ``The Fermi Surface" (Clarendon,
Oxford, 1973), pp.66-70.
\smallskip
\item{[50]}S.T. Chui, R.V. Kasowski, and W.Y. Hsu, Europhys. Lett. {\bf 9}, 385
(1989).
\smallskip
\item{[51]}R.J. Elliott, Phys. Rev. {\bf 96}, 280 (1954); L.M. Falicov and M.H.
Cohen, Phys. Rev. {\bf 130}, 92 (1963).
\smallskip
\item{[52]}G.F. Koster, Phys. Rev. {\bf 127}, 2044 (1962).
\bigskip
\bigskip
\centerline{\bf Figure Captions}
\bigskip
\item{Fig.~1}Fermi surfaces at vHs associated with (a) LTT phase; (b) LTO
phase.
\bigskip
\item{Fig.~2}LTO phase diagram based on strong umklapp scattering (Eq. 22).
Solid lines correspond to $J_0$ = 200, 300, 400, or 500K, in order of
increasing
gap, $D$.
\bigskip
\item{Fig.~3}Lowest energy eigenvalues of Mathieu's equation, Eq. 32a.  Both
the
exact solutions (solid lines) and the approximate solutions of Eq. 36d ($E_+$ =
dashed line; $E_-$ = dotted line) are shown.
\bigskip
\item{Fig.~4}Solid lines = potential well of Eq. 39, $\hat E\equiv 2H_{tilt}/
\tilde\alpha_-^{e\prime}\bar R^2$, with $<cos\phi >$ = 0.2, for (a) LTT and (b)
LTO phases.  Dashed lines are approximations to the potential using Eqs. 31 +
38b.
\bigskip
\item{Fig.~5}Eigenfunctions of Eq. 39 for the LTT (dashed lines) and LTO phases
(solid lines), assuming $\bar H$ = 400K, $H_0^{\prime}$ = -200K, for several
values of the overlap $S$ = 0.001 (a), 0.01 (b), 0.1 (c), and 0.5 (d).
\bigskip
\item{Fig.~6}Graphical solution of Eq. 42, where Arg is the right hand side of
Eq. 42 (solid lines) and the dashed line is $<cos\phi >$.  The solution of Eq.
42 is at the point where solid and dashed lines intersect (filled circles).
The calculations assume the same parameters as Fig. 5, with $S=0.01$, and the
various solid lines correspond to $T$ = 100, 80, 60, 40, or 20K, in order of
increasing $<cos\phi >$.
\bigskip
\item{Fig.~7}Equilibrium values of $<cos\phi >$ (solutions of Eq. 42) for the
LTT (dashed lines) and LTO phases (solid lines).  The parameters of Eq. 42 were
chosen as $\bar H$ = 400K, $H_0^{\prime}$ = -200K, with variable $S$. In order
of decreasing values of $<cos\phi >$ or $T_C$, the values of $S$ in part (a)
are 0.0001, 0.01, 0.02, 0.05, and 0.1.  In part (b), the values of $S$ are, in
order of {\it increasing} $T_C$, 0.3, 0.4, 0.48, 0.5, 0.52, and 0.55.  The
transition temperature $T_C$ has a minimum around $S\simeq 0.2$.
\bigskip
\item{Fig.~8}Free energies of LTT (dashed lines) and LTO phases (solid lines),
for the same parameters as in Fig. 7.  Increasing $S$ corresponds to increasing
magnitude of $F$ in part (a), but decreasing magnitude of $F$ in part (b).
Filled circles in part (b) show LTO-LTT crossover.
\bigskip
\item{Fig.~9}Phase diagram of dynamic JT transitions.  Solid lines show
HTT$\rightarrow$LTO and LTO$\rightarrow$LTT phase boundaries, $T_c(S)$, for the
same parameters as Figs. 5-8: $\bar H=400K$, $H_0^{\prime}=-200K$.  Dashed
lines are for the alternative parameters $\bar H=600K$, $H_0^{\prime}=-900K$.
\bigskip
\item{Fig.~10}Eigenfunctions of Eq. I1, for 2$\times$2 (top) and 4$\times$4
(bottom) clusters.  In each cluster, only the Cu's are indicated.  The
$\pm$-sign corresponds to the relative phase of the Cu $d_{x^2-y^2}$ orbitals
(periodic boundary conditions assumed).
\bigskip
\item{Fig.~11}Partial densities-of-states for the four subbands corresponding
to Fig. 10a (dashed lines) along with the total dos (solid lines). Inset:
Brillouin zone showing subband boundaries.
\bigskip
\item{Fig.~12}Fermi surfaces of LTO phase, with orthorhombic distortion: (a)
ordinary cell, including spin-orbit coupling. Circles on zone boundary show
points where bands must be degenerate, filled circle implies degeneracy for
all points along $k_z$, open circle implies degeneracy for special points only.
(b) Double zone appropriate when spin-orbit coupling is absent.
\bigskip
\item{Fig.~13}Group theoretical analysis of orthorhombic Brillouin zone. (a)
Real space cell of orthorhombic phase, showing orthorhombic cell (long dashed
lines) and inscribed primitive cell (short dashed lines).  Dotted lines = glide
planes.  The open and filled circles
represent Cu atoms, with filled circles representing atoms at $x = 0, a^*$, and
open circles atoms at $x = a^*/2$.  The solid lines represent the canted planes
of the `planar' O's. (b) Full Brillouin zone, corresponding to primitive cell.
Dashed lines = inscribed pseudocell.  Light shading represents portions of
Brillouin zone surface where the bands are twofold degenerate in the absence of
spin-orbit coupling (i.e., the $X-L-N$ face and the line $N-M$).  Heavy shading
indicates those portions where the degeneracy persists in the presence of
spin-orbit coupling (the line $L-X$ and the point $M$). (c) Orthorhombic
pseudozone.  Shading has same meaning as in Fig. 13b. (d) Zone folding
construction of second pseudozone.  Here shading illustrates how sections are
folded from the full zone into the pseudozone.
\bigskip
\item{Fig.~14}Eigenvalues of fractional Mathieu's function.  Solid lines =
solutions to the integral Mathieu's function and boundaries of stability for
the fractional values.  Fractional solutions are shown for orders 1/3 (short
dashed lines), 1/4 (long dashed lines), and 1/5 (dotted lines). [After Ref.
[33].]
\end